\documentclass[a4paper,11pt]{article}
\usepackage{aaskaiid}
\usepackage{orcidlink}

\title{Exploring the Galactic plasma with pulsars in the SKA Era}
\ShortTitle{Pulsars explore the Galactic plasma with the SKA}

\author[1,2]{Caterina Tiburzi\,\orcidlink{https://orcid.org/0000-0001-6651-4811}}
\ShortName{C. Tiburzi et al.} 

\author[3,4,5]{M.~T.~Lam\,\orcidlink{https://orcid.org/0000-0003-0721-651X}}
\author[6,7]{D.~J.~Reardon\,\orcidlink{https://orcid.org/0000-0002-2035-4688}}
\author[8,9]{N.\,\orcidlink{}~K.~Porayko\,\orcidlink{https://orcid.org/0000-0002-9604-6753}}
\author[10]{M.~Mevius\,\orcidlink{https://orcid.org/0000-0002-6524-9830}}
\author[11,12]{S.~K.~Ocker\,\orcidlink{https://orcid.org/0000-0002-4941-5333}}
\author[13]{S.~C.~Susarla\,\orcidlink{https://orcid.org/0000-0003-4332-8201}}
\author[14,15]{J.~R.~Dawson\,\orcidlink{https://orcid.org/0000-0003-0235-3347}}
\author[16]{J.~P.~W.~Verbiest\,\orcidlink{https://orcid.org/0000-0002-4088-896X}}
\author[6,7]{A.~Deller\,\orcidlink{https://orcid.org/0000-0001-9434-3837}}
\author[17,18,1]{G.~M.~Shaifullah\,\orcidlink{https://orcid.org/0000-0002-8452-4834}}
\author[19]{N.~D.~R.~Bhat\,\orcidlink{https://orcid.org/0000-0002-8383-5059}}
\author[20,21]{J.-M.~Grie{\ss}meier\,\orcidlink{https://orcid.org/0000-0003-3362-7996}}
\author[22]{M.~Walker\,\orcidlink{https://orcid.org/0000-0002-5603-3982}}
\author[23,24]{W.~Jing\,\orcidlink{https://orcid.org/0000-0002-1056-5895}}
\author[25,1]{F.~A.~Iraci\,\orcidlink{https://orcid.org/0009-0001-0068-4727}}
\author[26]{M.~Geyer\,\orcidlink{https://orcid.org/0000-0002-2822-1919}}
\author[27]{L.~Levin\,\orcidlink{https://orcid.org/0000-0002-2034-2986}}
\author[27]{M.~J.~Keith\,\orcidlink{https://orcid.org/0000-0001-5567-5492}}
\author[]{The SKA Pulsar Science Working Group}

\affiliation[1]{INAF-Osservatorio Astronomico di Cagliari, Via Della Scienza 5, I-09047 Selargius, Italy}
\affiliation[2]{INAF-Osservatorio Astronomico di Brera, Via Emilio Bianchi 46, I-23807 Merate, Italy}
\affiliation[3]{SETI Institute, 339 N Bernardo Ave Suite 200, Mountain View, CA 94043, USA}
\affiliation[4]{School of Physics and Astronomy, Rochester Institute of Technology, Rochester, NY 14623, USA}
\affiliation[5]{Laboratory for Multiwavelength Astrophysics, Rochester Institute of Technology, Rochester, NY 14623, USA}
\affiliation[6]{Centre for Astrophysics and Supercomputing, Swinburne University of Technology, PO Box 218, Hawthorn, VIC 3122, Australia}
\affiliation[7]{Australian Research Council Centre of Excellence for Gravitational Wave Discovery (OzGrav)}
\affiliation[8]{Sternberg Astronomical Institute, Moscow State University, Universitetsky pr., 13, Moscow 119234, Russia}
\affiliation[9]{Max-Planck-Institut für Radioastronomie, Auf dem Hügel 69, 53121 Bonn, Germany}
\affiliation[10]{ASTRON, Netherlands Institute for Radio Astronomy, Oude Hoogeveensedijk 4, 7991 PD, Dwingeloo, The Netherlands}
\affiliation[11]{Cahill Center for Astronomy and Astrophysics, California Institute of Technology, Pasadena, CA 91125, USA}
\affiliation[12]{Observatories of the Carnegie Institution for Science, Pasadena, CA 91101, USA}
\affiliation[13]{Physics, School of Natural Sciences, Ollscoil na Gaillimhe -- University of Galway, University Road, Galway, H91 TK33, Ireland}
\affiliation[14]{School of Mathematical and Physical Sciences and Astrophysics and Space Technologies Research Centre, Macquarie University, NSW 2109, Australia}
\affiliation[15]{CSIRO Astronomy \& Space Science, Australia Telescope National Facility, P.O. Box 76, Epping, NSW 1710, Australia}
\affiliation[16]{Florida Space Institute, University of Central Florida, 12354 Research Parkway, Orlando, FL 32826, USA}
\affiliation[17]{Dipartimento di Fisica “G. Occhialini”, Università degli Studi di Milano-Bicocca, Piazza della Scienza 3, 20126 Milano, Italy}
\affiliation[18]{INFN, Sezione di Milano-Bicocca, Piazza della Scienza 3, I-20126 Milano, Italy}
\affiliation[19]{International Centre for Radio Astronomy Research, Curtin University, Bentley, WA 6102, Australia}

\affiliation[20]{LPC2E, OSUC, Univ Orleans, CNRS, CNES, Observatoire de Paris, F-45071 Orleans, France}
\affiliation[21]{ORN, Observatoire de Paris, Universit\'{e} PSL, Univ Orl\'{e}ans, CNRS, 18330 Nan\c{c}ay, France}
\affiliation[22]{Manly Astrophysics, 15/41-42 East Esplanade, Manly, NSW 2095, Australia}
\affiliation[23]{National Astronomical Observatories, Chinese Academy of Sciences, Beijing, China}
\affiliation[24]{School of Astronomy and Space Sciences, University of Chinese Academy of Sciences, Beijing, China}
\affiliation[25]{Dipartimento di Fisica, Università di Cagliari, Cittadella Universitaria, I-09042 Monserrato (CA), Italy}

\affiliation[26]{High Energy Physics, Cosmology \& Astrophysics Theory (HEPCAT) Group, Department of Mathematics and Applied Mathematics, University of Cape Town,  Rondebosch 7701, South Africa}
\affiliation[27]{Jodrell Bank Centre for Astrophysics, Department of Physics and Astronomy, The University of Manchester, Manchester M13 9PL, UK}

\abstract{The ionised media that permeate the Milky Way have been active topics of research since the discovery of pulsars in 1967. 
In fact, pulsars allow one to study several aspects of said plasma, such as their column density, turbulence, scattering measures, and discrete, intervening structures between the neutron star and the observer, and aspects of the magnetic field throughout. Such sources of information allow us to characterise the electron distribution in the terrestrial ionosphere, the Solar Wind, and our Galaxy and have an important impact on other experiments involving pulsars such as Pulsar Timing Arrays. In this article, we review the state-of-the-art of plasma research using pulsars, the aspects that should be taken into consideration for optimal plasma studies, and we provide future perspectives on improvements to those enabled by the SKA.}

\begin{document}
\maketitle

\section{Introduction}
The broadband radio emission from pulsars travels through the Galactic interstellar medium (ISM), the Solar Wind (SW), and the ionosphere before reaching Earth, where it is collected by radio telescopes that can reveal it to an observer. Along this journey, the radiation of a pulsar undergoes deformations because of the ionised fraction of the aforementioned media. These variations speak of several characteristics of the ionised medium, such as the column density of free electrons, the turbulence of the medium, etc., and therefore they became new, powerful methods for studying plasma since the discovery of pulsars in 1967.
The precision and accuracy with which the observables related to such propagation effects were calculated have been significantly refined over the years, thanks to a progressive improvement in the observing facilities in terms of sensitivity, bandwidth, and radio-frequency coverage, and computing power. The advent of the Square Kilometer Array (SKA) – and in particular of SKA AA4, mid and low – will represent a milestone in each of these aspects and will be a turning point in our understanding of the Galactic, heliospheric, and ionospheric environments. The new findings will lead to a clear description of the electron density distribution in the Milky Way, and this will strongly impact other pulsar and radio-transient fields; in particular, it will allow us to precisely calculate the distance to pulsars, subtract the Milky Way electron contribution to the extra-galactic Fast Radio Bursts \citep{bai22}, and, importantly, it will yield a strong handle on the noise induced by the ionised interstellar medium in high-precision experiments such as Pulsar Timing Arrays \citep{vvp24}.
In this chapter, we explore several topics in the study of the ionised media through pulsars, and we compare the state-of-the-art with the expectations held for the dawn of the SKA Era.

\section{Dispersion measure and dispersion measure variations}
\label{sec:dmv}
The main effect induced by an ionised medium on an incident, broadband radio wave is \textit{dispersion}, i.e. the introduction of a frequency-dependent refractive index in the light propagation speed. In a cold plasma, this results in a frequency-dependent delay in the time-of-arrival (ToA) of the radiation (see e.g. \citealt{handbook}):

\begin{equation}
T_2 - T_1 =\frac{\mathrm{DM}}{K} \left ( \frac{1}{\nu_1^2} - \frac{1}{\nu_2^2} \right)
\end{equation}

where $T_1$ and $T_2$ are the ToAs of the incoming radiation (in seconds) with the observing frequencies $\nu_1$ and $\nu_2$ (in MHz), K is the dispersion constant (approximately $2.41 \times 10^{-4}$ cm$^{-3}$ pc MHz$^{-2}$ s$^{-1}$), and DM is the \textit{dispersion measure} parameter:

\begin{equation}
    \mathrm{DM}[\textrm{pc cm}^{-3}] = \int_{LoS} n_e(l) {\rm d}l
\end{equation}

with $n_e$ being the electron density (in cm$^{-3}$), LoS is the line-of-sight, and the distance to the pulsar is expressed in parsecs.

Pulsars are among the few objects where DM is measurable accurately, through the technique of pulsar timing (see e.g. \citealt{handbook}), and this opens up a variety of fields and applications spanning from the distance to pulsars \citep{sch12}, to the turbulence of the ionised medium \citep{ars95} and the reconstruction of the Galactic distribution of electrons \citep{ymw17}, and experiments such as Pulsar Timing Arrays (PTAs, see e.g. \citealt{vvp24} for a review on recent results and \citealt{Shannon01.2026.SKA} from this book).
In particular, the past decade witnessed a surge of studies around the topic of DM \textit{variations}, derived from long-term monitoring observations of pulsars that move transversally and radially and whose LoS spans large parts of the ionised medium in the Galaxy. Because such a medium (may it be the ionised ISM -- IISM, the SW, or the ionosphere) is turbulent, this motion induces fluctuations in the DM value over time. Those fluctuations inform us about the distribution, ordering, and kinematics of the ionised media on a number of spatial scales.
Moreover, this phenomenon affects high-precision pulsar timing experiments like PTAs by appearing as red, frequency-dependent \textit{noise} in the ToAs (see, e.g. \citealt{EPTADR2II, NG15noise, reardon2023} and \citealt{Shannon01.2026.SKA} from this book).

Finally, DMs can be supplemented by another observable resulting from propagation effects, namely the rotation measure (RM):
\begin{equation}
\textrm{RM}[\textrm{rad m}^{-2}]=0.81 \int_{\textrm{LoS}}n_e (\textbf{\textit{B}} \cdot \textrm{d}\textbf{\textit{l}}),
\label{eq:farad_rotat}
\end{equation}
which characterises the level at which the plane of linear polarisation rotates due to Faraday rotation, as a function of $n_e$ and of the magnetic field $\textbf{\textit{B}}$ (in $\mu\textrm{G}$) of the intervening media, integrated across the pulsar's distance expressed in parsecs. The projection of the magnetic field $\textbf{\textit{B}}$ on the displacement vector $\textrm{d}\textbf{\textit{l}}$ is expressed within the corresponding inner product. As can be seen from the expression above, RMs provide additional information about the magnetic component of the Galaxy. Therefore, combined DM and RM data serve as an excellent tool to reconstruct the large-scale structure of the magnetic field \citep{2006ApJ...642..868H}.

\subsection{Large-Scale Studies of DM Variations}
\citet{lam16} presented predictions for a wide range of DM variations, containing both stochastic (from the Earth-pulsar LoS passing through the turbulent IISM following a Kolmogorov regime) and systematic trends (from the changing LoS of the Earth and the pulsar across interstellar density gradients, the SW, the ionosphere, etc.). Linear trends can specifically also result from motions of the Earth and pulsar parallel to the LoS through a uniform medium, and transverse motions across gradients or slabs, e.g., from pulsar-induced bow shocks, are also explored. Stochastic variations act as a red-noise process with a spectral index related to the electron-density wavenumber spectral index.

As a matter of fact, several studies (discussed below) presented long timeseries of DM measurements following numerous pulsars, especially millisecond pulsars (MSP) from PTA experiments, where this DM `noise' is one of the main sources of disturbance, masking the primary signal of interest -- that from low-frequency gravitational waves (see e.g. \citealt{EPTADR2II,NG15noise}). 

For example, \citet{rhc16} derived the DM timeseries of 20 MSPs in the Parkes PTA (PPTA), with the
majority showing a significant linear DM trend. Four sources include an annual sinusoidal DM term
caused by the LoS to the pulsar tracing out a spiral pattern due to the motion of the Earth for
low-velocity pulsars. A follow-up work in the same collaboration, \citet{cpb23}, studied 35 MSPs of
the PPTA using the Ultra-Wide bandwidth, Low-frequency (UWL) receiver at the Parkes radio telescope
and reached a DM precision of $4\times10^{-6}$ pc/cm$^3$ for the best pulsar and a typical DM
precision around $2\times 10^{-4}$ pc/cm$^{3}$ in data taken since late 2018. This was possible
thanks to the increased bandwidth of the UWL (ranging from 700 MHz to 4 GHz), which increased the
fractional bandwidth (defined as the bandwidth divided by the centre frequencies) to 1.4.
Within the North American Nanohertz Observatory for Gravitational Waves (NanoGRAV) PTA collaboration, \citet{jml17} investigated DM variations in 37 MSPs using data from the NANOGrav nine-year dataset, covering frequencies from 300 MHz to 2.4 GHz. Significant fluctuations are identified in 33 of the observed pulsars, with timescales as short as a few weeks. DM structure functions are calculated for 15 pulsars, and they are shown to mainly align with a Kolmogorov turbulence spectrum.
Another work originating from the PTA community is presented in \citet{kmj21}, which used the upgraded Giant Metrewave RadioTelescope (GMRT) and its 400$-$500 MHz and 1360$-$1460 MHz observing bands to track four MSPs. The findings show that while the 400$-$MHz band gives good DM precision, an improvement of an order of magnitude can be achieved by combining both bands. An observation of the low-ecliptic MSP~J2145$-$0750 was found to be affected by a coronal mass ejection during the Solar transit.

A dramatic increase in DM sensitivity was achieved in the early 2010s using one of the European SKA pathfinders, the LOw Frequency ARray (LOFAR), which operates below a frequency of 240 MHz. \citet{dvt+20} analysed the DM variations in 36 MSPs observed with the LOFAR array over a timespan of about seven years (since late 2012) across the 100-200 MHz bandwidth, where they calculated a single DM value per observation (which differs from the aforementioned works where a number of observations taken at different epochs were averaged together to obtain a measurement of the column density). The amplitude of the observed DM variations ranges between $10^{-4}$ and $10^{-3}$ pc cm$^{-3}$ over several years, reflecting the influence of the IISM. The authors imply that potential linear trends here are a simple consequence of the limited timespan across which the IISM turbulence is observed. \citet{lam16} noted that such linear trends can sometimes be attributed to an increasing or decreasing Earth-pulsar distance or being characteristic of stochastic timeseries, and that by fitting and removing such linear trends it is possible to reduce the power in the power spectra of the DM fluctuations. \citet{dvt+20} also detected SW DM variations in 12 pulsars with angular distances from the ecliptic between $0.1^\circ$ and $38.8^\circ$. The results yielded by the DM structure-function analysis align well with a Kolmogorov turbulence spectrum, suggesting that the turbulence observed is consistent with theoretical expectations for interstellar turbulence. Overall, their work achieves a remarkably low median DM uncertainty of typically around $2\times 10^{-4}$ pc/cm$^{-3}$ for PTA MSPs and they find significant DM variations in all pulsars with such uncertainties or better. 

Partner to the LOFAR array is the French New Extension in Nançay Upgrading LOFAR (NenuFAR) interferometer \citep[][]{ZarkaNenufarIEEE,ZarkaNenuFARinstrument25}, which focusses on the $<$100 MHz band with increased sensitivity. \citet{bgt21} described the observing setup and the potential for pulsar science of the NenuFAR telescope. They can detect 12 MSPs below 100 MHz and show results for some slower pulsars down to as low as 16 MHz. In terms of IISM studies, they demonstrated the ability to observe dynamic spectra with clear diffractive scintillation (on PSR J0814+7429) and DM monitoring with extreme measurement precision of the order of $10^{-5}$ pc cm$^{-3}$ (on PSR J1921+2153), a much higher precision than the LOFAR 100$-$200 MHz band can obtain. The main challenge will be the detectability of pulsars at this low frequency, but NenuFAR is currently the most sensitive instrument in that frequency range, as its bandpass represents a clear improvement on the bandpass of LOFAR antennae below 100 MHz \citep[][]{ZarkaNenuFARinstrument25} and their collecting area (which was still growing at the time the paper was written) is much larger than that of the Long Wavelength Array (LWA)\footnote{see \url{https://nenufar.obs-nancay.fr/en/astronomer}}.

Taken jointly, the capabilities of LOFAR and NenuFAR have been used in \citet{iraci2025} to achieve the first inclusion of very-low-frequency data in a PTA dataset.
The study incorporates 12 LOFAR MSPs into the European PTA (EPTA) second data release, two of which also have complementary NenuFAR observations, resulting in an unprecedented frequency coverage spanning roughly 50 to 2500~MHz. 
This extended range significantly improves the constraints on the DM noise parameters and enhances the ability to disentangle DM variations from achromatic red-noise processes. 
Due to the high sensitivity of these low-frequency observations, the study also reveals limitations in the SW model used in PTA applications: it leaves a residual frequency-dependent signal that is subsequently absorbed into the DM modelling.
Overall, this work illustrated the impact that very-low-frequency data can have on PTA datasets and provides a clear indication of the role that such frequencies will play in the SKA era.

Using the SKA pathfinder MeerKAT under the ``Thousand Pulsar Array'' project, \citet{kj24} measured the DM time-series over a timespan of 4 years for 597 normal pulsars, resulting in 87 pulsars with a significant measurement of the DM gradient. Models of the IISM \citep{bhvf93} predict that the DM slopes, i.e. the DM derivative, should grow roughly with the square root of the DM, but \citet{kj24} found a steeper, mostly linear, dependence, consistent with recent results from IISM scattering \citep{kk2015}. Even when accounting for this relationship, the DM slopes for lower-DM pulsars (which include most MSPs) appear to be approximately an order of magnitude lower \citep{dvt+20,msb+23} than for pulsars with higher DM. \citet{kj24} attributed this to the greater velocity dispersion of the normal pulsar population, and hence the LoS traversing the IISM more quickly. 

 \subsection{Frequency-dependent Dispersion}
An additional consideration for wideband instruments with numerous frequency channels is that the calculation of a DM value might be affected by the \textit{chromatic DM} effect, a phenomenon that was theoretically described by \citet{css16} based on the fact that frequency-dependent multipath scattering is induced on the incident light rays by diffraction and refraction caused by electron density fluctuations on at least AU-scales. Therefore, the scattering of a ray bundle itself becomes frequency-dependent, and as each ray path differs and traverses a slightly different electron column density, the net effect is a frequency-dependent DM, after averaging over the light rays belonging to the same frequency. The root-mean-square (rms) of the DM difference between two specific frequencies, using a specific set of assumptions and a Kolmogorov spectrum, is found to depend on the inverse functions of the observing frequencies and on the square of either the Fresnel phase or the scattering measure. In theory, these assumptions result in DM variations that are smoother at low frequencies than at high frequencies.

The first detection of frequency-dependent interstellar dispersion was presented by \citet{dvt19}, based on LOFAR data spanning the 100$-$200 MHz frequency band, where they demonstrated that the DM timeseries for the upper and lower halves of their band differ significantly. They also detected pulse-shape changes due to the temporal evolution of scattering, but quantified the impact of this scattering on the DM timeseries and convincingly showed that the chromatic DM measurement is unaffected and suggested that the significant DM fluctuations might be caused by extreme scattering events. Although this work claimed that the frequency dependence of the DM variations is inconsistent with the expectations put forward by \citet{css16}; \citet{lld20} compared frequency-dependent DM due to ray path averaging over Kolmogorov turbulent structures with models of refraction from extreme scattering events and supported that the data for PSR J2219+4754 are in many aspects theoretically consistent with the former picture. The other results presented by \citet{don22} confirmed this mixed picture. Theoretical and observational research on frequency-dependent DM is hence an open field, far from being solved, and will continue to increase in significance as instruments grow in fractional bandwidth and sensitivity.

\subsection{Predictions for the SKA Era}\label{subsec:DM_exp}
The state-of-the-art has clearly evolved strongly over the past decade, with the advent of sensitive, low-frequency instruments with large fractional bandwidths and new PTA techniques. This has allowed a first glance at what the advent of the SKA-low and -mid, in their AA4, is bound to uncover, especially in the nascent topic of systematic DM variations.

The precision with which we can infer the DM values is, of course, critical for all of the pulsar
science cases. Without taking into consideration diffractive scintillation and variable scattering,
\citet{lbj14} derived an expression for the uncertainty of a single-epoch DM value, showing that in
the simplest case of a measurement based on two frequencies, the bottleneck is the precision of the
highest-frequency TOA. This implies that telescopes with a large collecting area are going to be
fundamental for DM estimates. At the same time, \citet{vs18} indicated the importance of the
fractional bandwidth for measurements of interstellar dispersion, also pointing towards wide-band instruments in addition to concentrating on the collecting area.

We can lay out an order-of-magnitude estimate of the sensitivity that will be offered by the AA4
configuration of SKA-low and -mid, based on the sensitivity curves reported in the
\textit{Anticipated SKA1 Science Performance}
document\footnote{\url{https://www.skao.int/sites/default/files/documents/SKAO-TEL-0000818-V2_SKA1_Science_Performance.pdf}}
(Tables 5 and 6 for, respectively, SKA-low and -mid) and by following parts of the formalism
summarised in \citet{lmc18}. For the sake of simplicity, we will ignore the impact of diffractive
interstellar scintillation, time-variable scattering, profile chromaticity\footnote{We note that
frequency-dependence of pulse profile shapes can substantially impact DM \emph{accuracy} by
introducing a constant offset from the true interstellar DM. However,
for most IISM studies and pulsar-timing analyses the accuracy of the measurement is not as important
as its precision and the ability to determine reliable variations, which is what this section
focuses on.}, and polarisation calibration errors (note that on the polarisation calibration errors of the SKA system it is not currently possible to have a handle yet). In determining these estimates, we bear in mind that, while the timing uncertainties usually come from the template-matching procedure, the high-precision MSPs currently observed with next-generation telescopes are chosen to be low-scattering sources and hence they will be largely dominated by pulse jitter.

Template-fitting contributes to a frequency-resolved ToA's uncertainty as $\sigma_{\mathrm{S/N}} (\nu) =  W_{\mathrm{eff}}/(S(\nu)\sqrt{N_{\phi}})$,
 where $W_{\mathrm{eff}}$ is the pulse's effective width, $S(\nu)$ is the signal-to-noise ratio (S/N) of the pulse peak, and $N_{\phi}$ is the number of phase bins. $S(\nu)$ can be calculated as $S(\nu) = \bar{S} (\nu) U (\nu) $, with $\bar{S} (\nu)$ being the period-averaged S/N and $U$ the inverse of the mean of the observed pulse shape when the peak is normalised to unity. $\bar{S}$ can be derived as $\bar{S} = I(\nu)\sqrt{N_{\mathrm{pol}}BT/N_{\phi}}/R$, with $I(\nu)$ being the frequency-resolved flux density ($I(\nu) = I_0  (\nu/\nu_0)^{-\alpha}$), while $\sqrt{N_{\mathrm{pol}}BT/N_{\phi}}$ represents the increase in flux given by a number of emission samples collected by using a certain bandwidth $B$, integration time $T$, $N_{\mathrm{pol}}$ polarisations and $N_{\phi}$ phase bins. Lastly, $R$ is the radiometer noise $R = T_{sys} (\nu)/A_e/ 2k= \left( [A_e/T_{sys}(\nu)]/(2760~\mathrm{m}^2/\mathrm{K})\right)^{-1}~\mathrm{Jy}$, with $T_{sys}$ and $A_e$ being, respectively, the system temperature and the effective area.  \\
On the other hand, the typical rms of single-pulse jitter is approximately 1\% of the pulse phase \citep{lcc16}, therefore, a MSP with a spin period of 2 ms will have a jitter contribution of $\sigma_{\mathrm{J,r}}\sim$20~$\mu$s per rotation, and in a 10-minute long observation (number of pulses $N_{\mathrm{p}}=3\times 10^5$), $\sigma_{\mathrm{J}}\sim \sigma_\mathrm{J,r}/\sqrt{N_{\rm p}}\sim20~{\mu}s/\sqrt{3\times10^5}\sim37~\mathrm{ns}$.\\
These two contributions to the ToA uncertainties are combined as a function of the frequency $\nu$ into a covariance matrix $C$, where jitter will cause non-diagonal elements to appear. The system can be solved with the least-squares formalism laid out by \citet{cs10} and subsequently re-described in \citet{lmc18}. Our basic timing model for the dispersive fit in matrix form can be written in the form

    \[
\begin{pmatrix}
    t_1\\
    t_{2}\\
    \vdots\\
    t_n
\end{pmatrix}
=
\begin{pmatrix}
    1 & K\nu_1^{-2}\\
    1 & K\nu_2^{-2} \\
    \vdots & \vdots \\
    1 & K\nu_n^{-2}
\end{pmatrix}
\begin{pmatrix}
    t_{\infty}\\
    \mathrm{DM}
\end{pmatrix}
+ 
\begin{pmatrix}
    \epsilon_1\\
    \epsilon_2\\
    \vdots\\
    \epsilon_n
\end{pmatrix}
\]

where the left-hand side is a vector containing the ToAs at frequencies $\nu_i$, and on the right-hand side is the design matrix $X$ (with $K$ being the dispersion constant), the parameter vector and the vector containing the errors described by the covariance matrix aforementioned. The second element of the covariance matrix of the parameters $( X^T C^{-1} X)^{-1}$, is the variance in the DM parameter.
By assuming the $A_{e}/T_{\mathrm{sys}}$ ratio predicted for SKA-low and -mid in the AA4 configuration, and a typical set of parameters for a MSP\footnote{A spin period of 2~ms, 500~$\mu$s of $W_{\mathrm{eff}}$, 1~mJy in flux density at 1.4 GHz with a spectral index of 1.6, 2 polarisations and 2048 phase bins, 1 MHz of channel width, an integration time of 10 minutes, and a scaling factor $U$ of 20, alongside an expected jitter rms in the order of 1\% of the pulse period.}, we reach expected DM uncertainties in the order of 10$^{-6}$ pc~cm$^{-3}$ for SKA-mid, and an exceptional 10$^{-8}$ pc~cm$^{-3}$ for SKA-low. This implies an improvement in the DM uncertainties by at least an order of magnitude compared to the best radiotelescopes currently available with SKA-mid in the AA4 configuration, and a clear competition with the ionosphere-induced rms \citep{Susarla24}. In the case of SKA-low, the uncertainties will definitely be lower than the rms imposed by the highly changeable ionosphere, which will hence introduce a significant scattering contribution to the DM timeseries. Therefore, these will be an asset for ionospheric modelling studies, as detailed in section~\ref{iono}.

\section{Scintillation observations and Galactic applications}

Interstellar scintillation of compact radio sources results from the scattering and subsequent interference of radio waves by plasma structures in the IISM of our Galaxy. Scintillation is observed in the dynamic spectrum as a modulation of the source flux density as a function of observing frequency and time. The dynamic spectrum is often analysed using its autocovariance function (ACF) to recover quantities such as the scintillation bandwidth and timescale, or, particularly in recent years, the ``secondary spectrum'' (power spectrum of the dynamic spectrum) to detect parabolic scintillation arcs that clearly demonstrate the geometry/kinematics of the scattering. 

Despite significant observational and theoretical progress, the compact plasma structures (AU-scale and smaller)  responsible for scintillation remain poorly understood. However, it has become clear that pulsar scintillation measurements not only provide a means to study this obscure plasma and the physics that governs it, but also the pulsars themselves. In binary pulsars, scintillation analyses have enabled accurate measurements of orbital parameters that are complementary to those from pulsar timing: scintillation provides information about motion in the plane of the sky, whereas pulse arrival times inform us about motion along the line-of-sight (LoS).

The recent focus on secondary spectrum analysis has come about because scintillation arcs have
enabled: precise LoS localisation of scattering material; measurements of the transverse motion of
pulsars and plasma; and has revealed that scattering can be highly anisotropic. Furthermore, recent
results \citep{Ocker+24,Reardon+20} suggest that the dominant source of interstellar scattering is plasma, which occupies only a tiny fraction of the LoS. Furthermore, studies of scintillating quasars have demonstrated that the plasma is very limited in its transverse extent, so the emerging picture is that the major part of the scattering material is made up of a vast number of tiny, but highly structured plasma regions.

The utility of scintillation has encouraged the development of open-source codes and new data representations. There is a growing synergy between pulsar timing and interstellar scintillometry, positioning these techniques at the forefront of precision pulsar astrophysics.

\subsection{Analysis Techniques}
Scintillation studies have required a variety of creative analysis methods to infer the properties of the source and scattering material. Although much work has focused on scintillation arcs, improved models of the two-dimensional ACF \citep{Rickett+14} allow for the inference of the degree of anisotropy as well as phase gradients, which can be integrated in time to recover an estimate of DM variations due to a scattering screen \citep{Reardon+23}

The curvature of scintillation arcs can be determined from the secondary spectrum via a Hough transform \citep{Bhat+16} or its variants, which represent the power along a scintillation arc as a function of the curvature \citep{Reardon+20}; this approach is widely applicable and can handle contributions from multiple plasma screens and instances where the scattering anisotropy is not particularly large. In the particular case of highly anisotropic scattering, data can be analysed with the transform $\theta$–$\theta$, which re-maps the secondary spectrum into angular coordinates assuming a one-dimensional scattered image and enables high precision arc curvature measurements \citep{Sprenger+21}. 
The incorporation of interferometric information offers an alternate pathway to reconstructing the stationary phase points that define the scattering screen based on secondary spectra that are formed from interferometer visibilities and hence encode geometric information in the secondary spectrum phase \cite[e.g.][]{Brisken2010}.

Holographic methods are new to pulsar astronomy, so the full impact of these techniques has not yet been felt. However, a wavefield representation of pulsar signals is useful for many applications because the electric field is simply a linear combination (a sum) of the many scattered waves --- in contrast to the field intensity (e.g. the dynamic spectrum), which is second-order in the field, or the secondary spectrum, which is fourth-order. Of the holographic methods under development, two require special conditions to be met: (i) phase retrieval applied to a dynamic spectrum requires the scattered signal to be sparse in some representation \citep{2008MNRAS.388.1214W, 2022MNRAS.510.4573B, oslowski+23}; and (ii) direct measurement of the (electric field) pulse broadening function is possible, but only for those few pulsars that emit giant (im)pulses \citep{Main+17, Mahajan+23}. Cyclic spectroscopy, on the other hand, is an intrinsically holographic method that can be applied to voltage data for any radio pulsar \citep{Demorest11}, although it produces higher phase-noise in the case of pulsars that are fainter and have longer periods. The analysis of cyclic spectra yields not only a measure of the impulse response function of the interstellar medium but also the intrinsic pulse profile of the pulsar, as would be seen if there were no interstellar scattering \citep{Demorest11, 2013ApJ...779...99W}. Cyclic spectra also have simultaneously high temporal and spectral resolution --- beyond the ``uncertainty principle'' limits of conventional spectroscopy \citep{Demorest11, Turner+24}. 

\subsection{The structure of the ionised interstellar medium}
Scintillation studies are uniquely positioned to enable studies of compact plasma structures in the
IISM. The IISM is turbulent, which drives a cascade of density fluctuations over a wide range of
spatial scales, leading to diffractive and refractive scintillation of compact radio
sources. However, the IISM is also filled with discrete structures that may dominate the scattering
and scintillation along many lines of sight. These can include extreme scattering events
\citep[inferred from pulsar scintillation bandwidths and DM variations,][]{Coles+15}, corrugated
reconnection current sheets \citep{Pen+14, Simard+18}, noodles or filaments \citep{Gwinn+19}, or
ionised skins of tiny cold hydrogen clumps \citep[suggested to explain extreme intra-day variability of some quasars,][]{ww98,Walker+17}.

The scintillation of pulsars can occasionally reveal scattering by their local environments \citep{Ocker+24}, allowing these local plasma structures to be studied. This includes within supernova remnants \citep{Yao+21}, pulsar bow shocks \citep{Reardon+25}, the interior of the Local Bubble \citep{Reardon+25}, and the wall of the Local Bubble \citep{Liu+23}. The scintillation of multiple sources across the sky can be connected to infer larger scale structures such as long plasma filaments in the Galaxy \citep{Wang+21}, although this requires a high density of sources on the sky and to date has only been possible with quasars.

\subsection{Scintillation surveys}

Large-scale observing campaigns show that scintillation arcs are present in almost all observations with a sufficient S/N and appropriate observation characteristics (e.g., wide bandwidth, fine channel resolution, and long integration time). A survey of 22 pulsars with DM $<100\,{\rm pc\,cm^{-3}}$ revealed scintillation arcs for 19 pulsars, implying that most nearby lines of sight can reveal one or two thin scattering screens \citep{Stinebring+22}. The morphology of the arcs can reveal the scattering geometry, with the presence of reverse arclets or a deep well of power along the conjugate frequency axis (the delay axis) of the secondary spectrum, suggesting highly anisotropic scattering. A LOFAR survey of 31 pulsars detected scintillation arcs in nine pulsars and measured diffractive scintillation parameters for 15 \citep{Wu+22}, with the scintillation unresolved (scintillation bandwidth less than a channel bandwidth) for other pulsars. The MeerKAT Thousand Pulsar Array captured scintillation arcs in 107 pulsars, including five pulsars that showed multiple scintillation arcs that are attributed to distinct screens \citep{Main+23}. This indicates that scintillation arcs are indeed ubiquitous, with their detection depending on S/N and observation characteristics.

Until recently, most pulsar secondary spectra showed just one scintillation arc, suggesting that the scintillation was dominated by one “thin screen” scattering region. Now, with higher S/N observations, multiple scattering screens towards individual sources have been inferred through the observation of multiple scintillation arcs, from six in PSR~J1136+1551 \citep{McKee+22} to nine in PSR~J1932+1059 \citep{Ocker+24}, and up to 25 in the nearest millisecond pulsar J0437$-$4715 \citep{Reardon+25}. It is now clear that not only are scintillation arcs expected for all pulsars, but that there should be an abundance of plasma structures causing such arcs along each given LoS. 

Long time-baseline studies add another dimension to these surveys. Ten‑year European Pulsar Timing Array (EPTA) monitoring of 13 PTA pulsars tracked secular changes in scintillation bandwidth and timescale, linking abrupt scattering enhancements to dispersion‑measure events \citep{Liu+22}. Imaging surveys with Australian Square Kilometre Array Pathfinder (ASKAP) uncovered minute‑timescale variability in dozens of active galactic nuclei (AGN) and seven known pulsars \citep{Wang+23}. Collectively, these surveys demonstrate that interstellar scattering is patchy on au scales yet widespread across the Galaxy, with multiple discrete screens often found along a single LoS. Whether these structures can be studied depends on the sensitivity of the telescope and the observing strategy.

\subsection{Pulsar applications}

A particular focus of the last decade has been the application of scintillation to understanding binary pulsars. The variable scintillation timescales of two relativistic binaries, PSRs~J0737$-$3039A \citep{Rickett+14} and J1141$-$6545 \citep{Reardon+19}, provided measurements of orbital inclination and sky orientation, in addition to the properties of the IISM (such as screen distance and degree of anisotropy). 

The precision of these binary pulsar studies was enhanced by the analysis of their scintillation arcs. Variations in the curvature of scintillation arcs were detected for the binary millisecond pulsar J0437$-$4715 \citep{Reardon+20} using 16 years of observations from the Parkes Pulsar Timing Array (PPTA). Variations in arc curvature follow from changing orbital velocities of the pulsar and the Earth, and enable precision measurements of the inclination and orientation of the pulsar orbit that can rival the precision of pulsar timing \citep{Reardon+20}. Annual and orbital variations were also modelled for PSR~J1643$-$1224 using the Large European Array for Pulsars (LEAP) \citep{Mall+22}, and PSRs~J1603$-$7202 \citep[including scintillation arcs during an extreme scattering event,][]{Walker+22} and J1909$-$3744 \citep[one of the most precisely timed millisecond pulsars,][]{Askew+23} using data from the Parkes Pulsar Timing Array (PPTA).

Scintillation has a range of other utilities, including pulsar detection through variance images \citep{Dai+16}, resolving the pulsar emission regions \citep{Main+18}, inferring DM variations \citep{Reardon+19}, measuring interstellar delays (including annual arc curvature variations) over long time periods \citep{Main+20}, and demonstrating three-dimensional spin-velocity alignment \citep{Yao+21}.

\subsection{Expectations for the SKA Era}

\begin{figure*}
    \centering
    \includegraphics[width=0.47\linewidth, trim={0 0 60 0},clip]{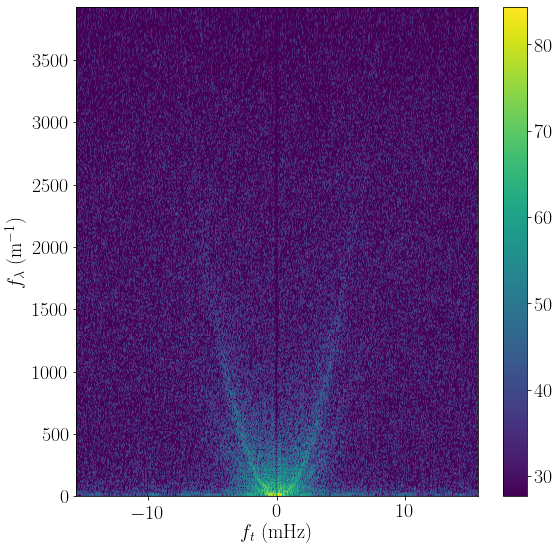}
    \includegraphics[width=0.45\linewidth]{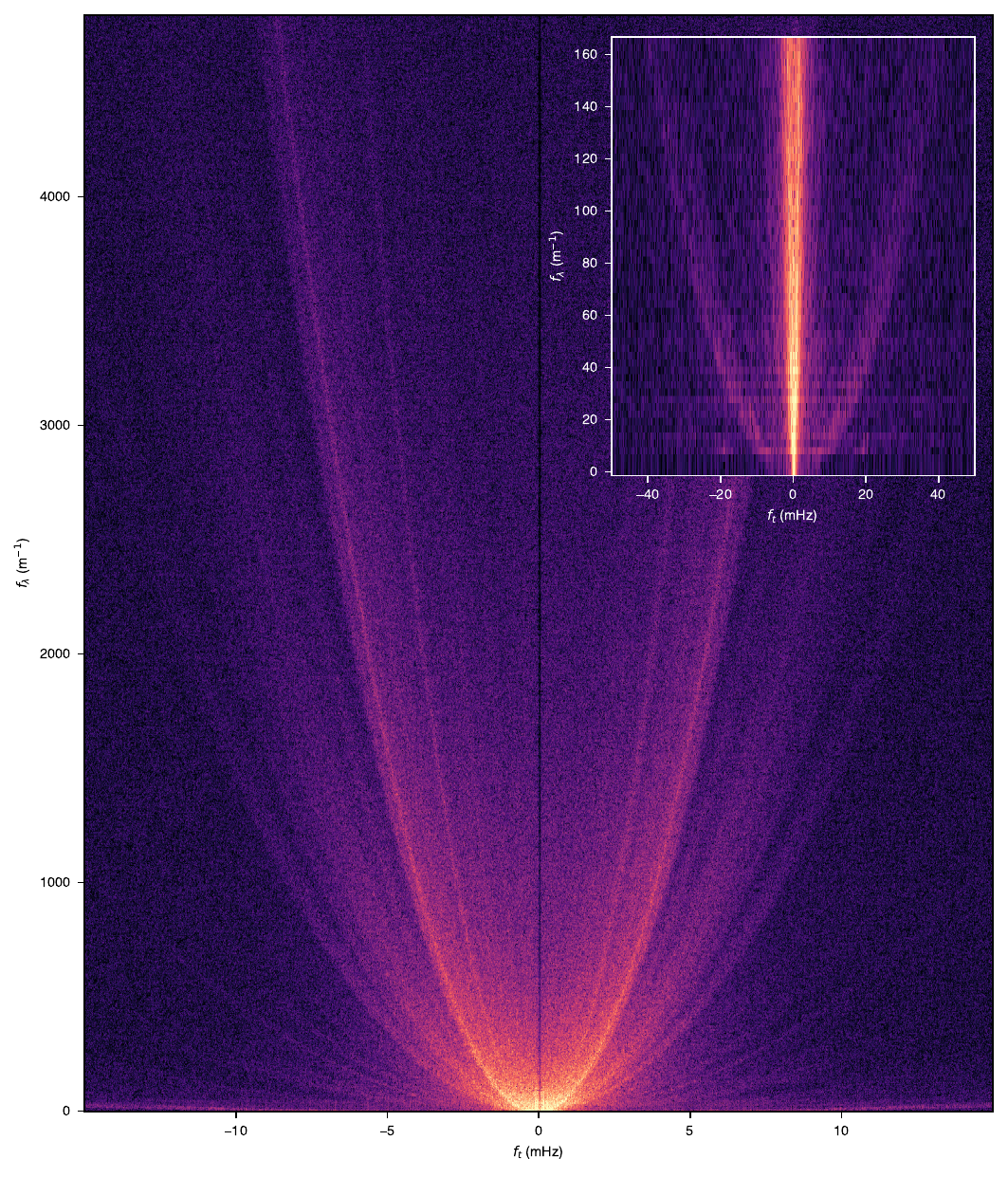}
    \caption{\textbf{Left:} Secondary spectra for PSR~J0437$-$4715 from long ($>10$ hour) observations with Murriyang, the 64-m Parkes radio telescope \citep[][]{Reardon+20}. \textbf{Right:} Secondary spectra for PSR~J0437$-$4715 from long ($>10$ hour) observations with MeerKAT radio telescope \citep[][]{Reardon+25}.}
    \label{fig:0437_sspec}
\end{figure*}

\begin{figure}
    \centering
    \includegraphics[trim=1.5cm 0cm 1.8cm 0.5cm,width=0.9\linewidth]{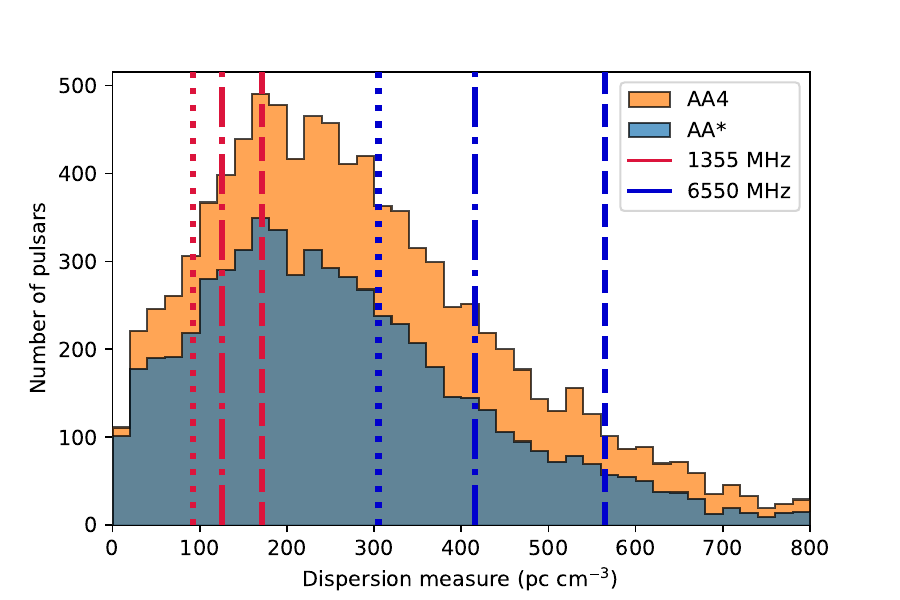}
    \caption{Distribution of dispersion measures for a simulated pulsar population observed with SKA-Mid configurations AA* (blue histogram) and AA4 (orange histogram). Vertical lines show the estimated maximum DM for which scintillation is resolved at frequencies corresponding to the centre frequencies of Band 2 (1355\,MHz; red) and Band 5a (6550\,MHz; blue). We have assumed three possible observing modes, with 1024 frequency channels (dotted), 4096 channels (dot dash), and 16384 channels (dashed) across the bands.}
    \label{fig:SKA-Mid-resolved}
\end{figure}

Scintillation observing programmes are often optimised by the choice of observing characteristics, for example, the resolution and length of the observations in frequency and time. In the SKA era, the S/N will be secondary to these characteristics for most pulsars that have resolved scintillation (e.g. low- to moderate-DM pulsars). For almost all measurements of diffractive scintillation scales, the uncertainty is dominated by finite scintle errors that cannot be overcome with a higher S/N. However, the S/N and quantity of scintillation arcs in secondary spectra can be greatly improved with greater telescope sensitivity because the arc morphology (e.g. degree of arc curvature) is geometric in origin and independent of the number of scintles. Figure \ref{fig:0437_sspec} shows the difference in secondary spectra obtained for the same millisecond pulsar, J0437$-$4715 from Murriyang (left panel) and MeerKAT (right panel) with comparable observation integration times. MeerKAT is approximately seven times more sensitive than Murriyang, and revealed 25 scintillation arcs, while only two were identified in the Murriyang observation. SKA-Mid AA4 is approximately four times the sensitivity of MeerKAT.

In order to predict how many pulsars will have resolved scintillation in a channelised observation, we estimate the scintillation bandwidth of a population of pulsars observable with the SKA-Mid AA* and AA4 configurations as predicted by \citealt{Keane01.2026.SKA} from this book, using the DM-scattering timescale relationship in \citet{coc22}. The scintillation bandwidth is the reciprocal of the scattering timescale and we assume that it scales with the observing frequency $\propto f^{4.4}$ according to the expectation of Kolmogorov turbulence. The assumed distribution of pulsar DMs is shown in the histograms in Figure \ref{fig:SKA-Mid-resolved}. 

We computed the maximum DM of a pulsar with resolved scintillation under three assumed observing modes (1024, 4096, and 16384 frequency channels) in two SKA-Mid bands (Band 2 and Band 5a). These are shown in vertical lines in Figure \ref{fig:SKA-Mid-resolved}. Scintillation is considered resolved at the centre frequency of the band if the scintillation bandwidth is more than a channel bandwidth. More than half of the observable pulsar population will have resolved scintillation from a 1024-channel observation with Band 5a (4600\,MHz to 8500\,MHz). SKA-Mid Band 2 (950\,MHz to 1760\,MHz) and lower frequencies (including SKA-Low) will be particularly useful for high S/N observations of low-DM pulsars. Finer channelisation through cyclic spectroscopy, or recording and offline processing of baseband data, may be required to resolve and study scintillation for pulsars with higher DMs at these lower frequencies.

\section{Pulse broadening and scattering measurements}

Measurements of pulse broadening times have long been used to steadily refine models for large-scale spatial distributions of the Galactic electron density and turbulence, such as the NE2001 model by \citet{cl02}. They have also provided useful information on the electron density spectrum and physics of turbulence, as well as for the empirical relation connecting the pulse-broadening time to the DM \citep[e.g.,][]{bcc2004}, which is widely used in pulsar population studies and survey simulations. The advent of new-generation low-frequency telescopes and wide-band instrumentation is poised to bring significant changes to the approach. In particular, new-generation telescopes are providing significant new insights, and have in particular enabled low-frequency detections ($\lesssim$300\,MHz) of hundreds of known pulsars using SKA-Low precursors and other facilities \citep{scb2019,bgt21,bsm2023,kts2025}, measurements for which can potentially provide important input for constraining the high-latitude distribution of electron density and the strength of turbulence, which are currently poorly constrained. The wide-band instrumentation available at sensitive telescopes such as the GMRT and MeerKAT can also enable measurements of frequency scaling indices for many more pulsar sight lines \citep[e.g.,][]{okp21} or scintillation measurements of low- to moderate-DM pulsars, both of which will highly complement low-frequency measurements. Furthermore, measurements of pulse broadening times over a large range in frequency (say, $\sim$1\,GHz to $\sim$100\,MHz), can also potentially allow the investigation of the frequency dependence of the pulse broadening function (PBF) -- a subtle effect theorised if inner scale effects are dominant.
 The next decade is thus poised for transformational improvements in our ability to characterise both the physics and distribution of the Galactic plasma turbulence. 

Despite the widespread use of electron density models for pulsar distance estimation such as NE2001 \citep{cl02} and YMW16 \citep{ymw17}, there are still large uncertainties in accurately predicting the degree of scattering, particularly at high Galactic latitudes. There are also notable discrepancies between the models and between the model predictions versus measurements. The successor of NE2001, currently under development, is expected to bring substantial improvements with its incorporation of a large number of H{\sc ii} regions. Constraints on angular broadening via VLBI can be combined with pulse broadening measurements to constrain where along the line of sight the scattering takes place, providing both tests of and the ability to improve models \citep[e.g.][]{Bower2014}. However, given the paucity of available pulse-broadening measurements at $|b|$  $\gtrsim$ $60^{\circ}$, the models are poorly constrained at higher-latitude sight lines, which are particularly important for interpreting measurements of Fast Radio Bursts (FRB). As more pulsar discoveries and measurements accrue for these high-latitude regions, e.g., through low-frequency scintillation and scattering measurements of nearby pulsars, we may expect further improvements to the local ISM models of \citet{occ20,occ2021}. Given the low to moderate DMs of many of these pulsars, low-frequency measurements, or wide-band scintillation observations using sensitive instruments such as MeerKAT \citep{Main+23}, are particularly relevant in this context. 

There have also been promising developments in the methods and techniques employed for the determination of pulse-broadening measurements and the impulse response function (IRF) of the IISM, i.e., the result of the propagation across the IISM of a delta-like signal. The latter is particularly useful for PTA applications, where there has been a growing interest in accurate modelling of the IISM noise budget in pulsar timing measurements. Although traditional methods such as forward modelling \citep[e.g.,][]{kk2015,gkk2017} are still in use, there has also been limited exploration of alternative methods such as deconvolution \citep{bcc2003,yl2024} and cyclic spectroscopy \citep[CS,][]{Demorest11}, which are promising for the characterisation of the IRF. With increasing access to baseband or voltage data and wide-band instrumentation, these new techniques are particularly promising and may allow us to accurately model the IISM noise budget in pulsar timing measurements. 
Early simulation work by \citet{psm2015} suggested that improved timing precision is likely achievable if CS-based correction can be routinely applied to PTA datasets. According to more recent work by \citet{dtj2021}, CS is most effective for highly scattered pulsars, if not limited by S/N, which may offer the possibility of doubling the PTA-quality pulsars.

The three methods are distinct, and they make a logical progression in their ability to robustly characterise the impulse response function (IRF) and the pulse broadening function (PBF, i.e., the broadening of a pulse profile); for instance, while the forward model approach involves the assumption of both the intrinsic pulse shape and PBF, the deconvolution approach involves no assumption of intrinsic pulse shape. CS, on the other hand, does not make either of the assumptions and thus, in principle, can be more robust in enabling the determination of IRF. With improved access to baseband or voltage data and more affordable real-time processing capabilities (for the routine application of CS), we may expect a surge of reliable measurements and IRF characterisation.  Given the typical high S/N values and impulsive nature of giant pulse emission, baseband descattering has been successfully applied to the giant pulses from e.g. PSR J1959+2048 \citep{Main+17}, directly probing the IRF. Improvements in IRF characterisation are particularly useful for improved IISM noise modelling in PTA datasets and ultimately for more unbiased characterisations of GW signals. 

PBF and IRF measurements not only allow for improved understanding of the IISM along the LoS and have impacts on PTA sensitivities but also can provide valuable insight into the pulsar environments and pulsar evolution by studying the electron densities and density gradients of plasmas surrounding pulsars. Typical examples include spider binary systems or pulsars embedded within supernova remnants. For example,~\citet{gsa+21} studied the distant Crab-twin pulsar, PSR J0540$-$6919 in the Large Magellanic Clouds (at 50 kpc) using MeerKAT, and found evidence for multiple scattering surfaces from analysing the rise-time of the giant pulse profile's leading edge, and the asymmetric scattering tail. Pulse profile rise-times could result from more than one scattering surface (and therefore more than one convolution with the IRF) each imparting a frequency-dependent phase-shift to the profile peak \citep[e.g.][]{jhw+25}. 

As per other applications, measured changes in the polarisation position angle observed across a pulse-broadening shape can also provide evidence for depolarisation from e.g. the nearby nebula and in future studies reveal RM-dependent travel paths. 
In the case of spider binary systems \citep{fst88}, observed changes in the PBF during eclipses directly probe the companion star's outflow material and thereby the nature and turbulence of such outflow winds. 

\subsection{Expectations for the SKA Era}
\label{subsec:PB_exp}
Currently, measurements of pulse broadening times are available for a modest fraction of known pulsars. In principle, these are obtainable from high-quality ($s/n\gtrsim 50$), high-fidelity pulse profiles in the frequency band optimal for making such measurements (i.e. $ w \lesssim \tau _d \lesssim P$, where $w$ is the pulse width and $P$ the period) and the measured pulse shape suffers minimal temporal smearing due to instrumental and other IISM effects (e.g. residual dispersion due to channelisation). The latter can be mitigated via coherent de-dispersion techniques. With an 1-2 order of magnitude improvement in sensitivity from AA4 and the impressive frequency coverage to be provided by the SKA-Low and Mid combinations ($\sim$50 MHz to $\sim$2 GHz), it will become feasible to obtain robust and accurate measurements  of $\tau _d$  for a large fraction of known pulsars (and optimistically for a good fraction of new ones to be uncovered by SKA searches). This will bring a major step increase in the sample of measurements for IISM modelling and will provide major improvements in the modelling of the spatial distribution of IISM turbulence and refinements of the scattering models. 

The wide frequency coverage and large fractional bandwidths provided by the mid- and low-combination will also enable reliable measurements of frequency scaling indices of pulse broadening times, which offer direct probes of the physics of IISM turbulence and micro-structure of the scattering material. Although there have been several insightful theoretical developments in this area, 
their applications have so far been limited to a modest number of cases. 
The frequency scaling laws of the scattering observables are expected to be LoS dependent, whereas the DM scaling relation of $\tau _d$ may vary for different segments of the Galaxy. For instance, the widely used $ \tau _d - \mathrm{DM}$ relation is evidently skewed by measurements in the Galactic plane, and a quantitative assessment of its directional dependence (e.g. the plane versus off the plane) has not been possible thus far. The major improvements to be brought about by both the sky distribution and the sample size of the measurements will allow us to investigate this for different segments of the sky or for different segments of the Galactic plane \citep[e.g.][]{jhw+25}.
This will be impactful in a number of areas, including pulsar populations, survey simulations, and the interpretation of FRB observations. 

Perhaps the most significant in the SKA era will be a transformation in our ability to characterise the microstructure in the IISM and explore the physics at these small physical scales ($\sim 10^6 -  10^{12}$\,m). High-fidelity measurements of pulse profiles along with the unprecedented sensitivity of AA4 ($\sim$20-30 times higher compared to currently operational telescopes in the low and mid bands), and the application of new techniques for de-scattering and cyclic spectroscopy, will allow us to characterise the IRF and PBF for many sight lines, thus opening a new realm in the exploration of pulsars as probes of IISM. These techniques are under active development 
and in the SKA era we can look forward to them maturing. Along with affordable computing, this will allow us to determine two of the powerful observables, i.e. the PBF/IRF and their frequency scaling laws routinely in pulsar observations. This will have multi-fold benefits; e.g. a) reconstruct the intrinsic pulse shape, which is widely used for studying pulsar emission physics and properties; b) construct the phase screen and infer the underlying IISM structure, which has the potential for improved timing precision in PTAs; and c) probe the interstellar turbulence, physics and the structure on scales that are not easily attainable by other observational probes.

\section{Pulsar distances and models of the Galactic distribution of free electrons}
Of the now $\sim3500$ pulsars known, just 10\% have independent distance measurements, while even fewer ($<5\%$) have a model-independent distance measurement of good precision ($20\%$ or better). Galactic electron density models remain the method by which most of the pulsar distances are estimated. The two most widely used models, NE2001 \citep{cl02} and YMW16 \citep{ymw17} use different methodologies and make distance predictions that differ by more than 50\% for many lines of sight \citep{pfd21}. Since the publication of YMW16, the most recent model, the sample of precise pulsar distances has nearly doubled due to large astrometric VLBI programmes  \citep{dgb19, dds23}, ongoing pulsar timing (e.g., \citealt{sbf24}), Gaia astrometry \citep{jkc18}, and the discovery of new globular cluster pulsars \citep{pqm21,prf24}. 
For distant Galactic-plane pulsars, even moderately precise distances measured from HI absorption spectra are valuable for modelling \citep{fw90}. Such measurements represent an important scientific goal for high-sensitivity telescopes \citep{jhh+23}.
High-sensitivity pulsar surveys with MeerKAT and FAST have indicated that there is still significant discovery space for pulsars, including those with large DM, RM, and scattering that reside deep in the Galactic plane \citep{hww21,okp21,pkj23}. This growing sample of new pulsars and pulsar distances indicates key areas for improvement in Galactic models, as well as heralding the IISM science that will be achievable with a substantial sample of pulsar distances.

\begin{figure*}
    \centering
    \includegraphics[width=\linewidth]{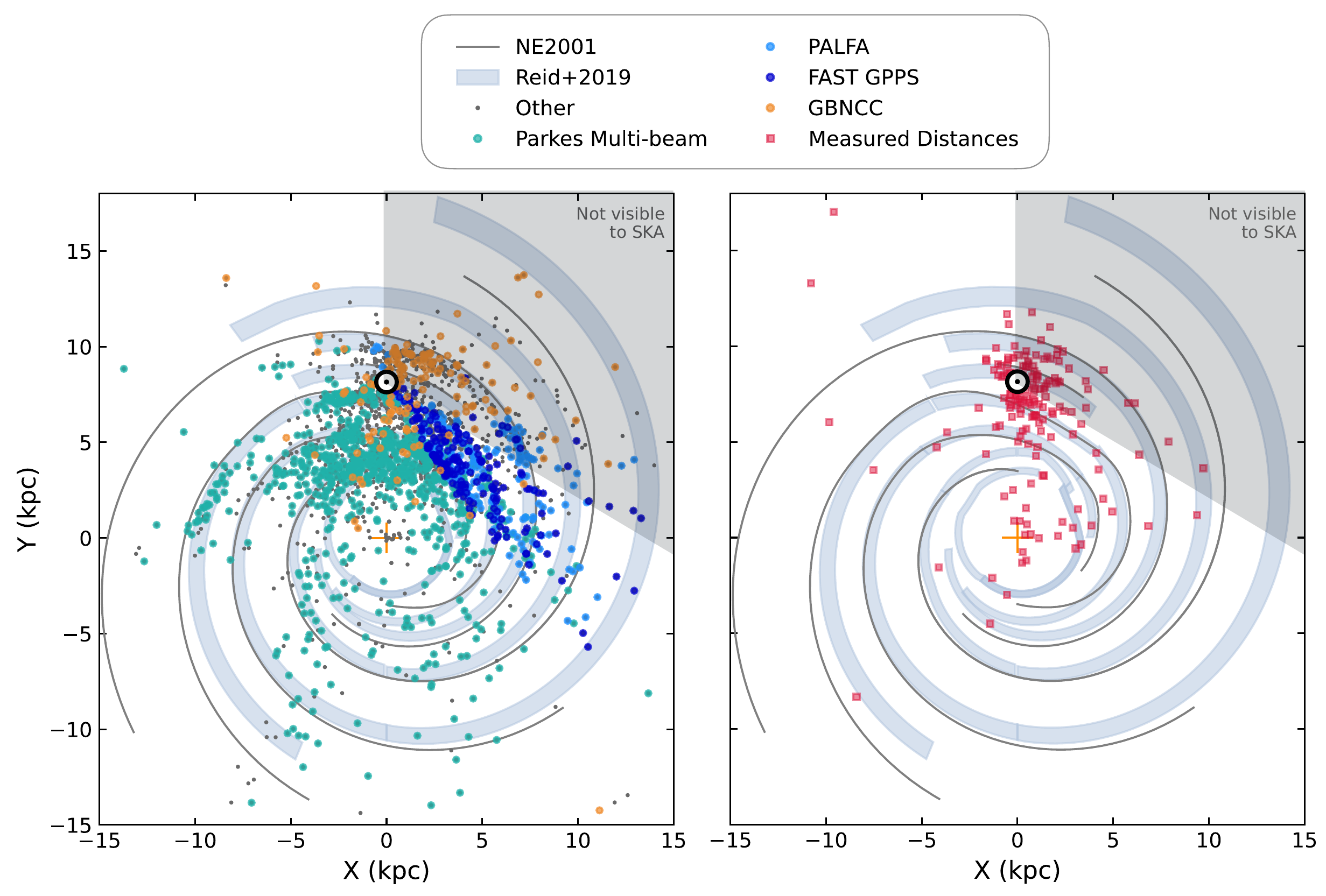}
    \caption{Distribution of known pulsars projected onto the Galactic plane, in galactocentric Cartesian coordinates. \textbf{Left:} Positions of all known radio pulsars, based on YMW16 distance estimates, with discoveries from four representative, major pulsar surveys highlighted: the Parkes multi-beam survey (teal), the Arecibo PALFA survey (light blue), the FAST Galactic Plane Pulsar survey (GPPS; dark blue), and the Green Bank North Celestial Cap survey (GBNCC; orange). \textbf{Right:} Pulsars with precise ($<25\%$ fractional uncertainty) distance measurements, based either on parallax or globular cluster associations. In both panels the Earth and Galactic centre are indicated by the open circle and cross, respectively; spiral arm models from NE2001 and \citealt{rmb19} are shown by the black and light blue shaded curves. Sky regions inaccessible to SKA are shown in grey.}
    \label{fig:psr-surveys}
\end{figure*}

It is increasingly evident that the Galactic electron density distribution is most poorly observationally constrained at low Galactic latitudes. Galactic plane pulsar surveys have revealed pulsars with DM and scattering larger than the rest of the known pulsar population, including pulsars with DM larger than the maximum predictions of NE2001 and YMW16 \citep{jkk20,okp21,hww21,pkj23}. Cross-matching known radio pulsars with comprehensive HII region catalogues suggests that HII regions may be intersected by as much as 30\% of known radio pulsar sightlines and that HII regions can account for a significant fraction of the total electron column density in the plane \citep{oal24}. For SKA surveys covering wider and deeper portions of the Galactic disc, HII regions and other types of discrete structure (e.g. supernova remnants and bubbles) will become increasingly relevant to the accurate interpretation of observed pulsar properties as tracers of the underlying electron density distribution. Current and ongoing multi-wavelength surveys of the ISM offer critical additional handles on the electron density contribution of both discrete structures and diffuse gas, e.g., via mapping of star-forming regions in spiral arms \citep{hh14,rmb19} and integral-field spectroscopy of diffuse gas \citep{dbk24}. 

\subsection{Expectations for the SKA Era}
The current sample of precise pulsar distances is heavily biassed by the declination coverage of the
Very Long Baseline Array (VLBA), leaving a dearth of distances below $\delta \lesssim
-25^\circ$. With its Southern all-sky coverage, SKA has the potential to significantly increase the
number of pulsar distances at Galactic longitudes $l < 0^\circ$, for which there is a critical gap
in the sample of measured pulsar distances (see \citealt{Keane01.2026.SKA} from this
book). Figure~\ref{fig:psr-surveys} shows the distribution of known pulsars projected onto the
Galactic plane, including all known radio pulsars and those pulsars that have precisely determined
($<25\%$ fractional uncertainty) distances. Despite there being comprehensive coverage of $l <
0^\circ$ in pulsar discovery surveys (due in large part to the Parkes multi-beam survey), the
majority of well-constrained pulsar distances are at $l>0^\circ$ and distances $<4$ kpc. The largest
distance measurements are currently based on globular cluster associations and are primarily at high
Galactic latitudes. Through a combination of pulsar timing and VLBI astrometry, SKA stands to
significantly improve our characterisation of the Galactic electron density distribution at
$l<0^\circ$, which is extremely sparsely sampled by current distance constraints, making any new
distance measurements (however few or imprecise) valuable in constraining Galactic electron-density
models. However, the strong scattering along many pulsar sightlines in this region will render many
sources unsuitable for SKA-VLBI astrometry until Bands 3 and/or 4 are added, where the angular
broadening will be too great in SKA-Mid Band 2 but the pulsars too faint in SKA-Mid Band
5.Consequently in the near future no large numbers of new SKA pulsar distances are anticipated.

\begin{figure*}
    \centering
    \includegraphics[width=\linewidth]{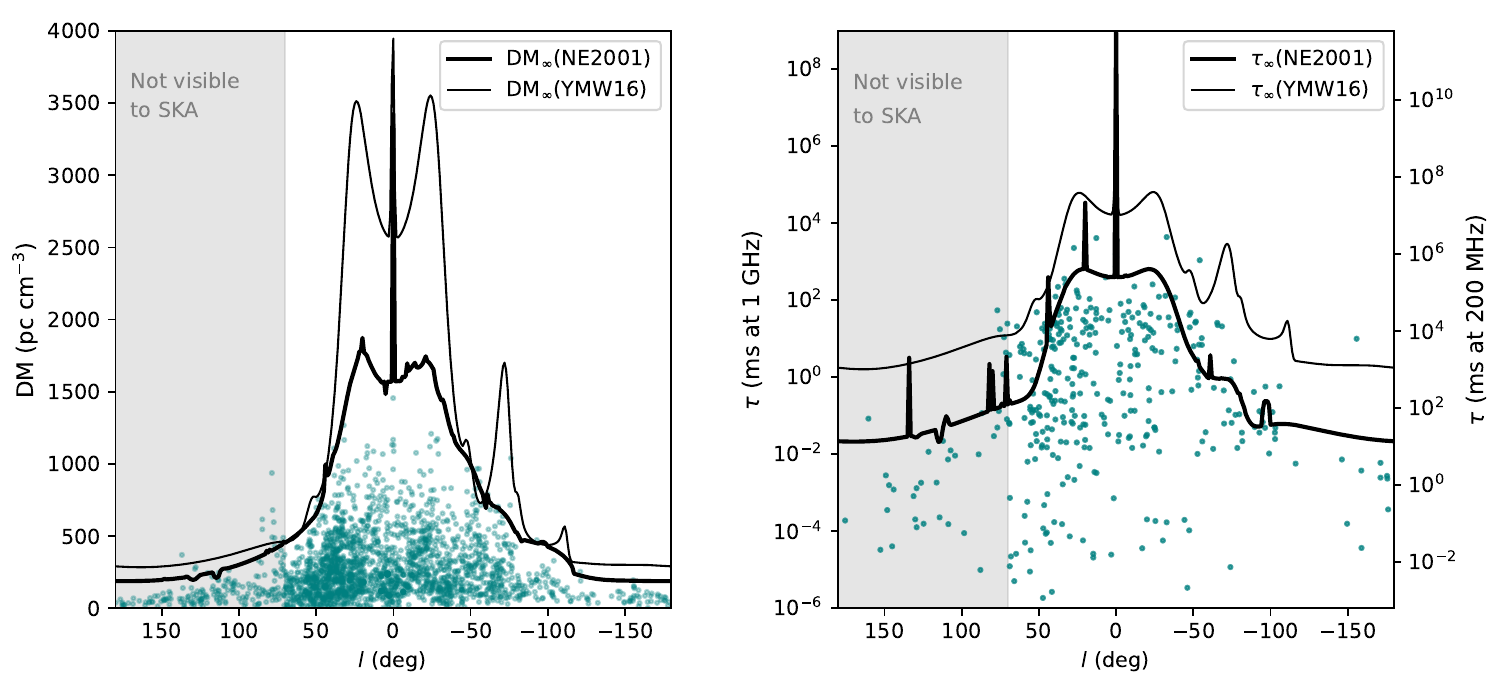}
    \caption{Comparison between observed DM and scattering distributions (teal points) and Galactic electron density model predictions (black curves) vs. Galactic longitude and for all available measurements at Galactic latitudes $|b|<10^\circ$. \textbf{Left:} DM vs. $l$ for known radio pulsars in the ATNF catalogue (teal), compared to the maximum DM predicted by NE2001 and YMW16 for sightlines integrated through the entire Galaxy at $b = 0^\circ$. \textbf{Right:} Scattering time ($\tau$) vs. $l$, based on the scattering time database compiled by \citealt{coc22}, compared to the maximum $\tau$ for NE2001 and YMW16 at $b=0^\circ$. Scattering times are scaled from 1 GHz to 200 MHz assuming $\tau \propto \nu^{-4}$. 
    }
    \label{fig:galmodel-comparisons}
\end{figure*}

Even in the absence of distance measurements, the factor $\sim2$ increase in known pulsars with the
SKA AA* configuration (and potentially up to $\sim5\times$ increase with AA4, see
\citealt{Keane01.2026.SKA} in this book), can be expected to constrain Galactic electron-density models in
regions of the Galaxy that were not fully covered by previous pulsar surveys. Specifically,
Figure~\ref{fig:galmodel-comparisons} shows how the observed DM and scattering distributions of
radio pulsars compare with the predictions of the most common electron density models. Especially in
the inner Galaxy, at
small and negative Galactic longitudes ($|l|\lesssim30^{\circ}$), both NE2001 and YMW16 predict
substantially higher maximum DMs than currently observed. Expanding pulsar discovery numbers in
that region of the Galaxy will therefore enable meaningful model constraints regardless whether the
distance to these pulsars is known or not.

There is one important caveat to these considerations, however. Specifically, nearly $50\%$ of
observed scattering timescales at $|l|<30^{\circ}$ reach $\tau \gtrsim 100$ ms at 200 MHz. Model
predictions of the yet-unknown pulsar populations in the inner Galaxy are even worse: the NE2001
model predicts scattering times of 1000 to 10,000 ms at 200 MHz for Galactic longitudes below
$30^{\circ}$, while the YMW16 model predicts scattering times up to two orders of magnitude higher
still (see again Figure~\ref{fig:galmodel-comparisons}), due to the combination of their higher
model DM in the inner Galaxy and their assumed empirical $\tau$-DM relation \citep{kk2015}. These
scattering times effectively prevent any pulsar discoveries in that region of the Galaxy with SKA-Low.

In contrast, the flux-density leverage from pulsars' steep ($\sim \nu^{-2}$) spectrum causes current predictions
\citep{Keane01.2026.SKA} to hold that the largest pulsar detection numbers will come from SKA-Low -- and
will therefore not contribute to the Galactic models in the areas where new pulsars are most
needed. In other words: SKA-discovered pulsars will still contribute meaningfully to Galactic
electron-density models, even if their distances cannot (yet) be measured; but mostly the pulsars
discovered in the inner Galaxy with SKA-Mid will prove useful, while the large number of predicted
SKA-Low discoveries will likely have a smaller impact on electron-density models, likely primarily
at higher latitudes.

Constraints on the Galactic electron density at high latitudes stand to benefit from combining globular cluster pulsar distances (the largest distances available), an increasing sample of pulsar parallaxes at distances greater than 1 kpc \citep{dgb19}, and low-DM FRBs that place upper bounds on the electron density content of the Galactic halo \citep{pz19,yt20,bgk21,cbg23}. The expanded sample of pulsar distances indicates that the electron density of pulsars relatively near the Solar system is well described by
a plane-parallel medium with a scale height of 1.6 kpc and a mid-plane density of 0.015 cm$^{-3}$ \citep{occ20}. Although the known population of Galactic halo pulsars remains small \citep{rcl18,xhw22}, combining future SKA-discovered halo pulsars with FRBs in the Local Volume likely offers the best prospects for directly constraining the halo electron density. Concurrently, expanding the sample of pulsars in the Large and Small Magellanic Clouds to higher DMs will probe deeper into these satellite galaxies, as demonstrated by searches with MeerKAT \citep{plg24}, and may ultimately constrain not only the electron density of these satellite galaxies but also their impact on the surrounding circumgalactic medium.

\section{Heliospheric measurements}

The heliosphere delineates the region surrounding the Sun and the solar system, with the SW being its primary content. The SW is a highly magnetised stream of plasma flowing out from the Sun as a result of the pressure of the hot solar corona. Understanding the dynamics of the heliosphere requires careful investigation of the electron and proton content and the embedded magnetic fields. The SW can have a direct bearing on various aspects of life on Earth; therefore, it is crucial to understand its characteristics, to explore its physical drivers as well as their impact on near-Earth space and other astrophysical experiments. Studying the SW can also give us a deeper understanding of the heliosphere and solar corona. Although such studies have been carried out using various spacecraft \citep{
ulysses, kooi2014}, spacecraft are a more financially demanding approach than ground-based measurements; and are typically limited to a few spatial configurations. A complementary approach to understand SW dynamics is to estimate the electron content \& magnetic field structure of the SW by studying the DMs and RMs of background compact radio sources such as pulsars \citep{counselman, You2007, tiburzi2019, madison2019, wood2020, Susarla24} and extragalactic sources \citep{hewish1964, coles1995, tokumaru2006, bisi2010, kooi2014, fallows2023, jensen2025}. 

In pulsar timing, the approach has traditionally been to model the SW as a time-independent,
spherically symmetric distribution of free electrons, but various recent studies (see below) have
highlighted the need for a time-variable model, particularly in the context of pulsar timing. In the
past ten years, we have seen significant advances in terms of understanding the heliosphere using
pulsar measurements. Particularly, the low-frequency observations from interferometers like LOFAR
have provided us with a huge leap towards precise measurements of dispersion measures and Faraday
rotation. Effectively, this has allowed pulsar-based heliospheric measurements to evolve from being
focused mainly on correcting the SW impact on high-precision timing data towards providing actual
contributions to heliophysics studies of the Solar magnetosphere and SW, as briefly sketched below.

\citet{You2007} proposed a bimodal SW model that accounted for fast and slow streams through distinct radial electron density distributions. Using solar magnetograms from the Wilcox Observatory, they attributed LoS components to individual streams and combined them to calculate the total SW contribution. However, \citet{tiburzi2019} tested both the bimodal and spherical SW models using low-frequency observations of PSR~J0034$-$0534 and found that neither fully explained the data, while the spherical model performed slightly better. This discrepancy with \citet{You2007} was attributed to either improved DM precision at lower frequencies or differences in the heliospheric latitude of the observed pulsars. Subsequently, \citet{ord2007} and \citet{You2012} discussed how measurements of the Faraday rotation of pulsars occulted by the SW can provide a unique opportunity to estimate the magnetic field near the Sun. This experiment had already been carried out earlier \citep{bird1980} in a limited fashion but has not been successfully reproduced with the more sensitive, modern, low-frequency telescopes.

Based on these findings, \citet{madison2019} analysed the NANOGrav 11-year dataset with a specific focus on the potential variations in the SW density. Although they did not find any evidence for density variations during the solar cycle, \citet{Tiburzi2021} did demonstrate that SW models with time-variable amplitude provide a more accurate description of the dispersive effects derived from highly sensitive LOFAR data.

Further advances were made by \citet{Hazboun2022}, who relaxed the traditional assumption of $1/r^2$ electron density in favour of a $1/r^\gamma$ profile where $r$ is the distance between the observatory and the Sun, and also compared binned and Fourier-based models for time-dependent SW variability, achieving better results for PTA datasets. Most recently, \citet{luliana2024} employed Gaussian processes to capture SW variability during solar conjunctions. Although their approach effectively modelled piecewise variability, their annual $n_{\rm e}$ fits failed to account for the continuous SW density fluctuations observed by spacecraft within the inner solar system. To improve on this model, \citet{Susarla24} employed a Bayesian approach using a continuously varying Gaussian process to model the SW in the LOFAR data. Their analysis reveals a strong correlation between the inferred electron density at 1 AU under a spherically-symmetric model and the ecliptic latitude (ELAT) of the pulsar. Pulsars with $|ELAT|<3^{\circ}$ exhibit significantly higher average electron densities. They also demonstrate the electron density of pulsars whose $|ELAT|>3^{\circ}$ correlates with the solar activity cycle. Although they demonstrate improved results, understanding the SW is a largely unanswered puzzle. With an improved sensitivity of SKA-Low, we can achieve much better estimates of the SW parameters, paving the way for a deeper understanding of space-weather dynamics.

\subsection{Expectations for the SKA Era}
The forthcoming SKA telescope, particularly its low-frequency component SKA-Low, promises to be a transformative instrument for probing the solar wind using pulsars. With its anticipated ability to measure DMs with extraordinary precision — reaching 10$^{-8}$ pc cm$^{-3}$ — SKA-Low will enable highly sensitive tracking of DM variations induced by dynamic solar activity, such as solar flares and Coronal Mass Ejections (CMEs). These abrupt events can lead to localised increases in the electron column density along the LoS to a pulsar, and the sensitivity of the SKA would allow for the detection of such variations with an unprecedented precision.

Assuming a simplified spherical model of the heliosphere, it is expected that column densities could be constrained to within 0.05\% accuracy. This would effectively allow for the creation of snapshots of the SW distribution at any given time just by observing pulsars. Although reconstructing a full three-dimensional model of the SW from these measurements remains a complex inverse problem, the enhanced spatial and temporal resolution offered by SKA may make this more feasible than ever before.

Moreover, by complementing DM measurements with RM analyses, SKA will provide insights not just into the density structure but also into the magnetic-field component of the solar wind. In past studies like \citet{hwd16}, pulsar observations have been used to estimate magnetic field strengths associated with CMEs by comparing radio propagation effects with white-light coronagraph observations. With the advanced polarimetric capabilities of the SKA, such measurements could become routine, significantly improving our understanding of the magnetised structure of solar transients.

\section{Ionospheric measurements}\label{iono}

In the last decade there has been a number of works on ionospheric studies using pulsar observables. However, despite its importance, the ionosphere still remains a rather niche topic in pulsar astronomy. Terrestrial plasma manifests itself most prominently in \textit{variations} of the Faraday rotation of pulsars, and dominates other contributions from astrophysical sources by many orders of magnitude \citep{ol2012}. The ionosphere poses a serious obstacle to probing the magneto-ionic media of the Universe. It also seriously degrades the sensitivity of pulsar polarization arrays, which perform searches for ultra-light axions, manifesting signals in pulsars' Stokes Q and U \citep{2025PhRvD.111f2005P, 2026PhRvL.136a1001X}. Building on the pioneering study by \cite{ssh2013, pnt2019, pmh2023} confirmed that a single layer model (SLM)\footnote{The SLM has been implemented in the publicly available RMextract (\url{https://github.com/lofar-astron/RMextract.git}) and ionFR (\url{https://github.com/csobey/ionFR.git}) software packages.} of the ionosphere, in which a terrestrial plasma is approximated as an infinitely thin slab fixed at a certain height, is not enough to describe the complex physics with sufficient accuracy. In particular, \cite{pnt2019} using LOFAR observations of four highly polarised sources, found that two types of systematics after subtracting the SLM remain: long-term linear positive trends correlated with the Solar cycle, and diurnal/annual quasi-periodicity\footnote{Given that a pulsar is observed approximately at the same zenith angle during the year, the daily variations are propagated to seasonal (with a period of 1 year) periodicities, so that diurnal and annual peaks are strongly covariant.}, reflecting the highly dynamic properties of the terrestrial plasma. In \cite{pmh2023} a step has been taken towards using more sophisticated ionospheric calibration of pulsar data. In particular, two models were considered: 1) thick layer model of the ionosphere, and 2) TOMographic IONosphere (TOMION) dual-layer voxel model \citep{hjs1997, hjs1999}. The first accounts for the thickness of the terrestrial plasma layer using the electron density distribution from the International Reference Ionosphere (IRI) \citep{br2008} scaled to the total electron content (TEC) values, obtained from the GPS-reconstructed ionospheric maps \citep[see also][]{gas2011}. The TOMION captures the dynamical behaviour of the ionosphere by using dual-layer voxel structure of the ionospheric layer. Although the considered models performed better than the SLM, none of them could completely remove the effect of the terrestrial plasma. There is currently an ongoing and promising study using an improved IRI-UP model \citep{ppr2018, ppr2018b} that takes into account the updated electron density profiles, to calibrate pulsar data \citep{ppa2026}. The idea of using the Faraday rotation of pulsars themselves to self-calibrate the data using advanced computational schemes, such as neural networks, will be explored in Usynina et al. (in prep.). In addition to pulsar observations, imaging provides a powerful tool to probe the \textit{differential} TEC of the ionosphere by measuring the differential phase delays between the elements of the interferometer \citep{mtp2016}. The conclusion is in line with the previous studies: the implementation of more complex modelling will be necessary \citep{amo2015, amo2016}. Therefore, the modelling of the ionosphere is still an open question, which should be explored by the combined efforts of radio astronomers and plasma physicists. Beam-forming and imaging data with SKA-Low will provide unprecedented sensitivity to the ionospheric Faraday rotation and avenues to investigate and improve our understanding of its physics.

\subsection{Expectations for the SKA Era}
\begin{figure}[h]
    \centering
	\includegraphics[width=\textwidth]{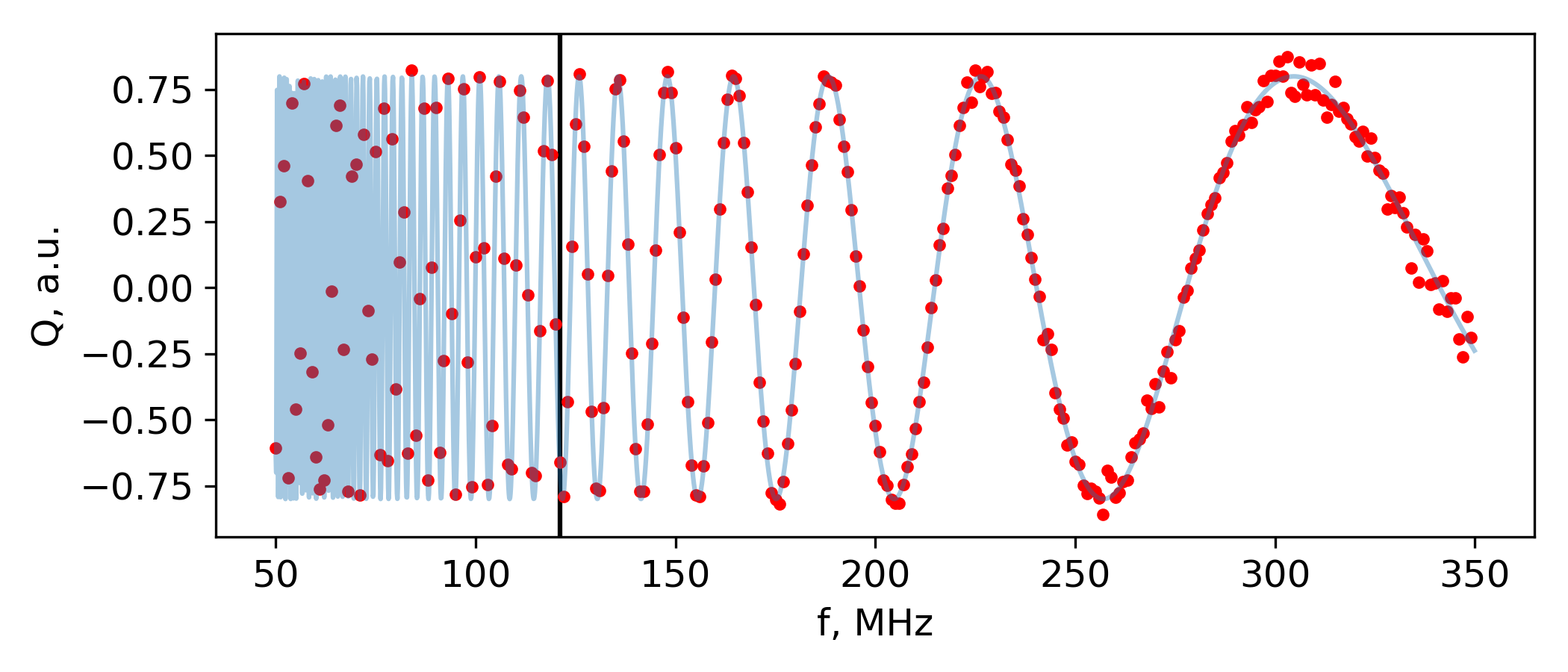}
    \caption{Effect of inter-channel depolarisation. oscillations of Stokes Q as a function of observing frequency for a source with RM=4~rad/m$^2$. The width of the frequency bin is 0.5~MHz. The effect of depolarisation is the most severe when there is less than one observing point per eighth of the oscillation period of Q. The zone of severe inter-channel polarisation is highlighted with the black line.}
    \label{fig:depol}
\end{figure}

\begin{figure}[h]
    \centering
	\includegraphics[width=\textwidth]{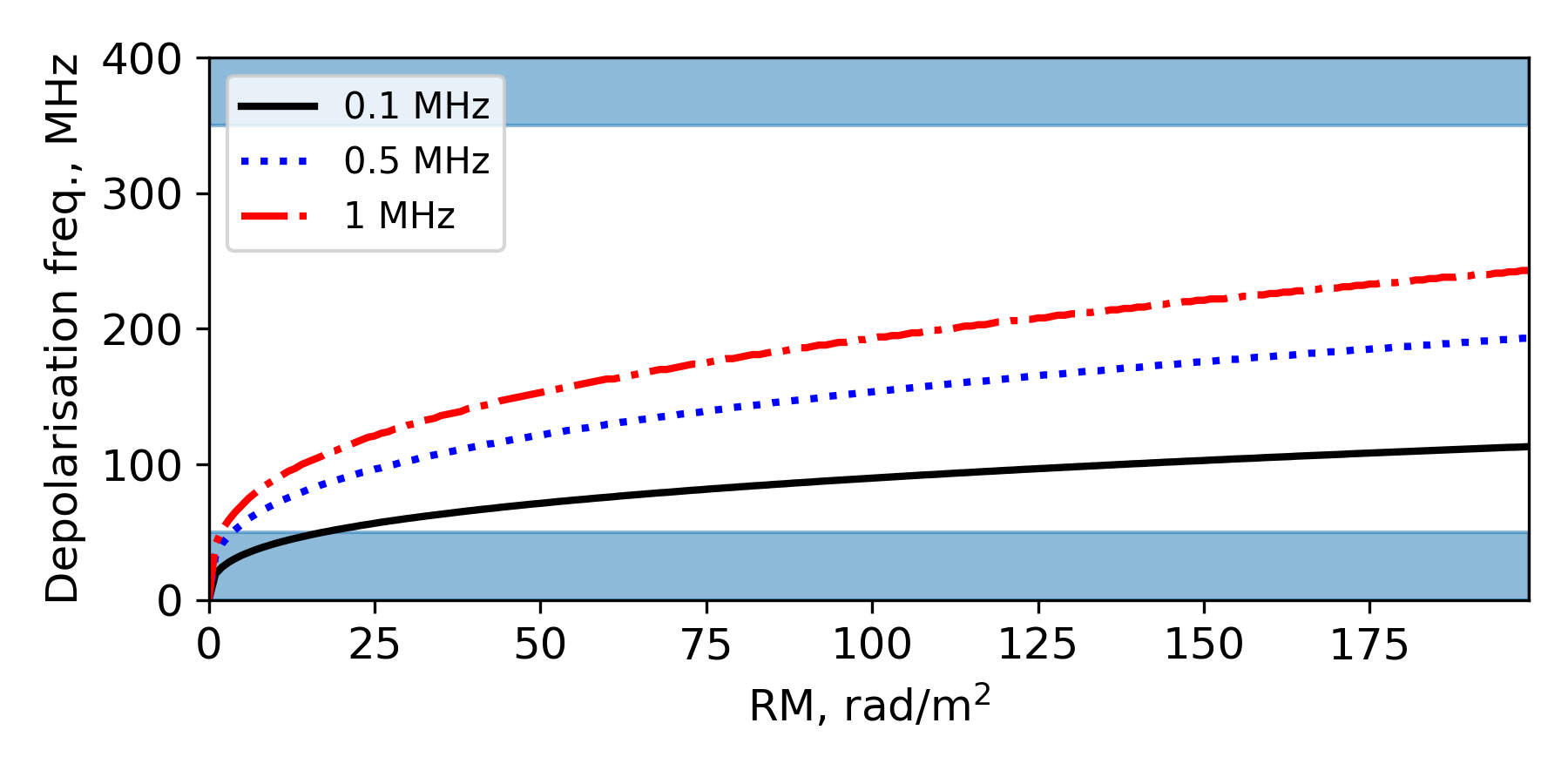}
    \caption{The effect of depolarisation is the most severe when there is less than one observing point per eighth of the oscillation period of Stokes Q/U. The plot shows this critical depolarisation frequency as a function of RM. Different lines demonstrate the magnitude of the effect for three channel widths. The not shaded region shows the frequency coverage of the SKA-Low.}
    \label{fig:depol_f}
\end{figure}

The terrestrial plasma will have an effect on the SKA measurements and will cause a serious
interfere with astrophysical observations. As one of the propagation effects, the Faraday rotation
of the ionosphere will have a particularly strong effect on observations at lower frequencies (see,
e.g., Figure~\ref{fig:depol}). Therefore, pulsar observation with the SKA-Low is the main focus of this subsection. For a pulsar with typical characteristics, described in Sec.~\ref{subsec:DM_exp}\footnote{Here we assume that the SKA-Low antennas with a frequency channel bandwidth of 1~MHz are used to observe a pulsar for 10 min.}, the expected precision at which the RM will be measured for the AA4 (and AA*) configuration is $\delta$RM$\sim10^{-4}$~rad/m$^2$.
However, this unprecedentedly high precision will be diminished by the severe \textit{inter-channel depolarisation}, which will affect the lower segment of the SKA-Low antennas. As was found by Bondonneau et al. (in prep.) with pulsar observations with the NenuFAR radio interferometer, it is practically impossible to fully recover the polarisation content of pulsar signal at the lower half of the NenuFAR band using classical \textit{incoherent}\footnote{Analogous to pulsar coherent and incoherent dedispersion.} Faraday de-rotation, i.e. when the correction for the Faraday rotation is performed after the data have been channelised. The inter-channel depolarisation is most significant in the channels, where Stokes Q(U) makes the full "turn", i.e. $\delta\phi=2\textrm{RM}(\lambda^2_i-\lambda^2_j)>2\pi$, where $\lambda_i$ and $\lambda_j$ are the central wavelengths of the two subsequent channels. Though, the effect is already noticeable, when $\delta\phi>\pi/4$. From expressions, it is clear that the accuracy degrades as the RM increases. Without coherent de-Faraday rotation the precision on the measured RM degrades by a factor 10$-$20 from the reported value for RM$\sim4$~rad/m$^2$. Coherent de-Faraday correction should be applied directly on the baseband data (before channelization), so that one needs to have a prior knowledge on the pulsar RM (including ionospheric contribution) with the precision of at least 1 rad/m$^2$ , which can be provided by, e.g., IRI model of the ionosphere. The effect of inter-channel depolarization for the SKA-Low is demonstrated in Fig.~\ref{fig:depol}. The critical frequency at which the effect of inter-channel depolarization beocmes prominent is depicted in Fig.~\ref{fig:depol_f}.

Even after applying coherent Faraday de-rotation to the data, the measured RMs cannot be directly used to access the astrophysical magnetic fields and electron content. The first step is to carefully subtract the contribution of the terrestrial plasma using existing models of the ionosphere. As was described in the previous section, the vast majority of these models heavily relies on the column electron densities monitored by the Global Navigation Satellite System (GNSS) infrastructure. 

\begin{figure}[h]
    \centering
	\includegraphics[width=\textwidth]{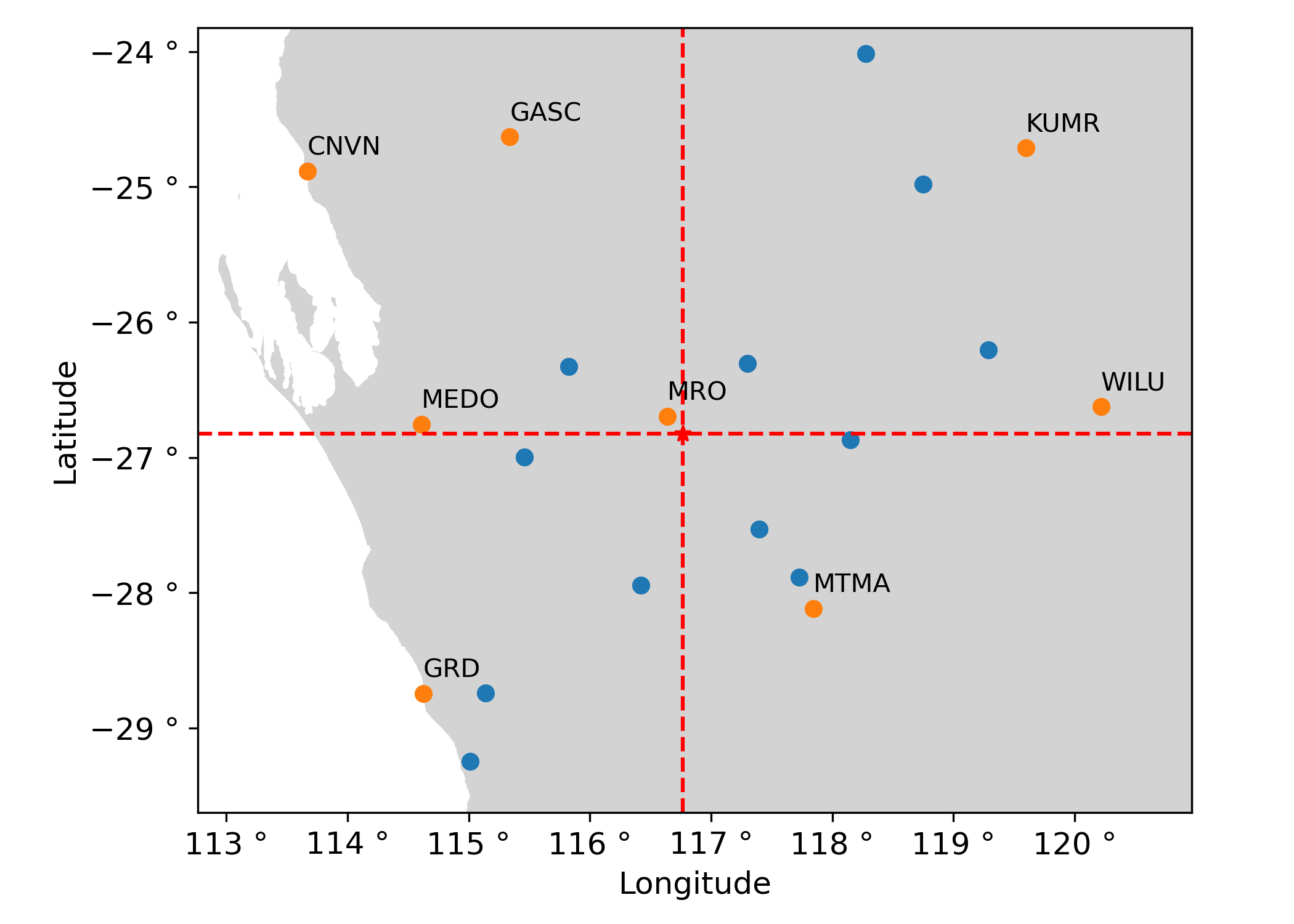}
    \caption{Current (orange) and proposed (blue) GNSS stations in the vicinity of the SKA-Low site (marked with red star). The total of 20 stations provide the precision to measured RM of $\sim0.02$ rad/m$^2$.}
    \label{fig:gps_cover}
\end{figure}

Given that the average electron density gradient is dTEC$\sim5\times10^{-3}$~TECu/km\footnote{\url{https://doi.org/10.5281/zenodo.15000430}}, in order to be able to measure RM with the precision of $\delta$RM$=10^{-4}~\textrm{rad/m}^2$ one needs to be able to probe the fluctuations in the ionospheric properties every $\delta d$:
\begin{equation}
\delta d=0.25\,\textrm{km}\left(\frac{\delta\textrm{RM}}{10^{-4}~\textrm{rad/m}^2}\right) \left(\frac{B_\textrm{ion}}{3\times10^{5}~\textrm{nT}}\right)^{-1} \left(\frac{\textrm{dTEC}}{5\times10^{-3}~\textrm{TECu/km}}\right)^{-1}\,.
\end{equation}
This implies that in order to successfully reconstruct the TEC gradients, there should be at least one LoS connecting GNSS station and a satellite that crosses the ionospheric layer every $0.25\times 0.25$~km$^2$. To estimate the optimal number of GNSS stations to provide the optimal coverage, we assume that both GPS (Global Positioning System) and GLONASS (GLObal NAvigation Satellite System) satellites are used for the tracking (48 satellites in total), resulting in $\sim10$ satellites visible simultaneously from the SKA-Low core site. In addition, we limit the altitude angle of pulsar observations to be above 30~deg (only core stations are used). This corresponds to more than $8\times10^5$ GNSS stations to be placed in the vicinity of the SKA-Low. This unfeasibly high number quickly reduces with the sensitivity that needs to be achieved. For example, for RM precision of 0.02~rad/m$^2$ one needs to have only 20 active GNSS stations around the core. An example of how the proposed stations should complement the existing ones are shown in Fig~\ref{fig:gps_cover}. Similar idea has been pointed out in \cite{amo2016}, who proposed to place 14 additional stations around MWA (Murchison Widefield Array) on the west coast of Australia. With such infrastructure one is able to resolve ionospheric structures with typical sizes of 10 - 100 km. One should note that this coverage configuration is not unique. More precise estimates on the exact positions of the GNSS stations around the SKA-Low will be performed in Khizriev et al. (in prep.). A system of strategically placed digisondes \citep{rg2011} will provide additional information on the vertical distribution of the electron density in the terrestrial plasma layer, and improve the quality of ionospheric correction.

Finally, even with poor GNSS coverage, having a subset of pulsars which will be observed on a regular basis with the RM precision of at least $\delta$RM$=10^{-3}~\textrm{rad/m}^2$, will potentially provide a unique information on ionospheric electron content at the typical scales of $\sim3$~km. Combined with the GNSS and ionosonde data, pulsar RMs will form a powerful tool to probe the ionosphere to unprecedent level. The capacity of this perspective methodology will be explored in future works.

\section{AU-scale fluctuations in HI absorption}

With a low ionisation fraction (typically $\ll$ 0.1), the atomic ISM is not commonly included in discussions of astrophysical plasmas. Nevertheless, pulsars have had a unique role in probing the structure and properties of the atomic medium -- most notably providing some of the only measurements of AU-scale fluctuations in HI absorption (sometimes termed Tiny Scale Atomic Structure; TSAS, see \citealt{heiles2007}). The observed fluctuations have characteristic spatial scales of 0.1$\sim$10,000 AU and are observed in cold neutral medium (CNM; $n\sim10$ cm$^{-3}$, $T\sim100$ K) 21cm absorption spectra \citep{stanimirovic18}. Variations are either spatial -- measured from maps of optical depth across resolved background sources; or temporal -- measured in pulsar absorption spectra as the pulsar moves behind a foreground cloud. The underlying principle is that the very small angular sizes and high proper motions of pulsars provide a pencil-thin absorption beam that traverses the foreground gas on scales of 10--100s of AU per year. \citep[e.g.][]{frail94,johnston03,minter05,stanimirovic10}. The data are folded at the period of the pulsar, and the off-pulse spectrum is subtracted from the on-pulse one to derive clean optical depth spectra that are sensitive only to the cold, optically thick component of the atomic medium. Combined with H{\sc i} emission spectra from the same sightline (simply the off-pulse spectrum in single dish studies, and off-pulse with zero spacing data added in interferometric work), the radiative transfer equations may be solved to obtain an estimate of the spin temperature, $T_\mathrm{s}$, and from that the gas kinetic temperature -- a critical quantity in estimating the gas pressure. When the pulsar's proper motion and cloud distance (and motion) are taken into account, this provides a set of measurements of $\Delta\tau$ (i.e., the time-variation in optical depth of the H{\sc i} line), $\Delta T_\mathrm{s}$ and $\Delta L$, the length scale, for every pair of epochs and for each detected H{\sc i} absorption component.

Although the existence of an AU-scale HI structure has been known since the 1970s \citep{dieter76}, the underlying physical processes remain poorly constrained. Under simple assumptions, observations imply cold ($\sim$ 100K), high-density ($n\sim10^{4-5}$ cm$^{-3}$) neutral cloudlets that are significantly overpressurised with respect to their surroundings \citep{stanimirovic18}. However, it seems likely that this ``overpressure problem'' will all-but vanish if the structures are significantly elongated along the LoS \citep{heiles97} -- a picture that is quite consistent with expected modes of cold gas formation via thermal instability in a magnetised medium \citep{inoue16}. Alternatively, it has been claimed that the observed opacity variations do not imply genuine density enhancements on the length scale of the optical depth variation, but can be  explained as arising from a superposition of turbulent structure on all scales \citep{deshpande00}. Such a structure should be ubiquitous and have a characteristic red power-law spectrum -- a picture supported by some VLBI studies \citep{roy12,dutta14}, and most recently by pulsar-based work \citep{liu25}. The `turbulence' and `discrete cloudlet' interpretations need not be mutually exclusive, particularly if we favour models in which neutral gas turbulence arises from the relative motions of many small, unresolved cold filaments and clumps embedded in a warmer medium \citep{koyama00,inoue12,inoue16}. 
In their 2018 review, \citet{stanimirovic18} collated all results from the previous literature (including a large number of non-detections) to favour a picture in which AU-scale fluctuations in HI absorption are a sporadic phenomenon, potentially arising from occasional local bursts of turbulent energy dissipation or local instability-driven fragmentation processes.
However, the parameter space sampled is still sparse and critical properties such as the characteristic distribution of size scales and optical depth fluctuations, the volume filling factor, and relationships between size and optical depth are poorly constrained. 

In the last 10 years, there have been only a handful of new observational works, and the physical picture remains uncertain. \citet{rybarczyk20} found absorption fluctuations to be almost ubiquitous, but probed only relatively large spatial scales ($\gtrsim$ 2000 AU) using background sources separated on the sky by orders of arcseconds. \citet{liu21} reported a single tentative detection against one pulsar (a 17 AU cloud against J1600$-$5044), and recently have performed densely-sampled multi-epoch absorption measurements against a different source (J1644$-$4559), finding possible evidence for a turbulent red power law spectrum in their single varying component. Most recently, \citet{jiang25} reported a tentative optical depth variation at 18 AU scales seen against the MSP J1939+2134, albeit with low certainty and low spectral resolution. 

\subsection{Expectations for the SKA Era}

The critical observational requirements for AU-scale HI absorption fluctuations are high optical depth sensitivity coupled with a high (epoch-to-epoch) sampling cadence. Reported $\Delta\tau$ range between $\sim$0.01--1.0 \citep{stanimirovic18}, but it is likely that the distribution skews towards low optical depths, particularly on the smaller spatial scales accessible to pulsar observations. The key advantage of the SKA is therefore primarily its large collecting area and outstanding sensitivity\footnote{Note that while the collecting area of FAST exceeds that of SKA AA4, pulsar H{\sc i} absorption studies with FAST must use the telescope in spectral line mode at its fastest dump rate of 1s (limiting the pool of available pulsars significantly and complicating data analysis), or record raw voltages (a mode generally not accessible to external users). In contrast, the SKA will be able to record high spectral resolution data folded to the period of a pulsar, affording significant advantages in terms of data volumes, access, and ease of processing.}. 
The optical depth sensitivity to H{\sc i} absorption may be expressed as $\sigma_{e^{-\tau}}(\nu) = [\sigma_{S_\mathrm{on}}^2(\nu) + \sigma_{S_\mathrm{off}}^2(\nu)]^{0.5} /\langle S_{\rm psr, on}\rangle$, where $\sigma_{S_\mathrm{on}}(\nu)$ and $\sigma_{S_\mathrm{off}}(\nu)$ are the spectral rms noises in the on-time-averaged pulse and off-pulse spectra, respectively, and $\langle S_{\rm psr,on}\rangle$ is the mean pulsar 20-cm flux density in the on-pulse phase. The spectral sensitivity terms come from the radiometer equation, with the usual dependence on $T_{\rm sys}/ (A_e \sqrt{\Delta\nu~t})$, where $t$ is the integration time and $\Delta\nu$ is the spectral channel width. 
For a given instrument, the pulsar flux density therefore has the largest impact on sensitivity. However, the target pool is limited to slow pulsars. This is because: (a) a bandwidth of at least 8\,MHz is normally required, to capture the full spread of Galactic velocities while leaving sufficient spectral baseline for calibration; (b) the raw frequency channel width should ideally be no more than $\sim$1 kHz ($\sim$0.2 km\,s$^{-1}$), to permit the resolution of narrow, cold H{\sc i} lines (though data can later be binned to improve S/N as desired). This limits the sampling interval to $\sim$1\,ms and the pool of viable sources to $P\gtrsim0.1$ s.  

For a new survey of AU-scale HI absorption fluctuations to surpass what has come before, we must move from the realm of individual sources into the realm of statistics. This requires a step-change in the number of detections. The SKA will allow for this by extending the available pool of viable background sources to lower flux density pulsars, while simultaneously achieving unprecedented sensitivities on the brightest. Excluding the very recent low confidence variation reported with FAST \citep{jiang25}, all previous pulsar-based detections of AU-scale HI absorption variations have been made against a mere 7 sources, all with $S_{1400}\gtrsim15$ mJy. With SKA AA4 we can push this limit down to $S_{1400}\sim5$ mJy -- increasing the viable source population by a factor of four -- and still be sensitive to optical depth fluctuations of $\Delta \tau\lesssim0.05$. For the brightest sources, our sensitivity will be over an order of magnitude better\footnote{For SKA AA4, in tied array mode including all baselines $<10$ km, $SEFD = 2kT_{\rm sys}/A_e = 2.1$ Jy. Assuming a sky temperature of 5\,K, an H{\sc i} peak brightness temperature of 100\,K, a total integration time of 5h, a pulse duty cycle of 0.1, and a binned channel width of 1\,km\,s$^{-1}$ gives $\sigma_{S_\mathrm{on}}(\nu)=0.6$--2.2\,mJy and $\sigma_{S_\mathrm{off}}(\nu)=0.2$--0.7\,mJy (where the range arises from the frequency-dependent variation in the H{\sc i} brightness temperature contribution to $T_{\rm sys}$). The equivalent range in optical depth sensitivity, $\sigma_{e^{-\tau}}(\nu)\approx\sigma_{\tau}(\nu)$, is $\sim$0.01--0.04 for $S_{1400}=5$\,mJy and $\sim$0.0006--0.002 for $S_{1400}=100$\,mJy.}.

\section{Conclusion}
In this chapter, we reviewed the main propagation effects affecting the travel of a pulsar's radiation across a number of Galactic plasma -- namely, the IISM, the SW and the terrestrial ionosphere -- and the observables that we can derive by studying the affected emission. In particular, we detailed the current state-of-the-art of research, focussing on the developments of the last ten years, and developed predictions on the improvements that will come from the advent of the Square Kilometer Array -mid and -low in their AA and AA4 configurations. Although clear breakthroughs are already expected with AA, AA4 is likely to be a game-changer in the near-future scientific panorama. The impacts of DM variations, scattering, scintillation and the Solar wind on experiments such as PTAs are reported in \citealt{Shannon01.2026.SKA} from this book.

\section*{Acknowledgments}
GMS acknowledges financial support from the European Union’s H2020 ERC Consolidator Grant “Binary Massive Black Hole Astrophysics” (B Massive, Grant Agreement: 818691) and the financial support provided under the European Union Advanced Grant “PINGU” (Grant Agreement: 101142079). MTL acknowledges support received from NSF AAG award number 2009468, and NSF Physics Frontiers Center award number 2020265, which supports the NANOGrav project. SKO is supported by the Brinson Foundation through the Brinson Prize Fellowship Program. DJR is partly funded through the ARC Centre of Excellence for Gravitational Wave Discovery (CE230100016). JPWV acknowledges support from the National Science Foundation (NSF) AccelNet award No.~2114721.

\bibliographystyle{abbrvnat-maxbibnames4}
\bibliography{./chapter,./aaskaii} 

\end{document}